\documentclass[fleqn,usenatbib]{mnras}

\usepackage{newtxtext,newtxmath}

\usepackage[T1]{fontenc}

\DeclareRobustCommand{\VAN}[3]{#2}
\let\VANthebibliography\thebibliography
\def\thebibliography{\DeclareRobustCommand{\VAN}[3]{##3}\VANthebibliography}

\usepackage{graphicx}	
\usepackage{amsmath}	
\usepackage[dvipsnames]{xcolor}
\definecolor{ao}{rgb}{0.0, 0.5, 0.0}

\title[Subhalo strong lensing tests]{Testing strong lensing subhalo detection with a cosmological simulation}
\author[Q. He et al.]{Qiuhan He$^{1}$\thanks{E-mail: qiuhan.he@durham.ac.uk},
James Nightingale$^{1, 2}$,
Andrew Robertson$^{3}$,
Aristeidis Amvrosiadis$^{1}$,
Shaun Cole$^{1}$,
\newauthor
Carlos S. Frenk$^{1}$,
Richard Massey$^{1, 2}$,
Ran Li$^{4, 5}$,
Nicola C. Amorisco$^{1}$,
R. Benton Metcalf$^{6}$,
Xiaoyue Cao$^{4, 5}$,
\newauthor
Amy Etherington$^{1, 2}$
\\
$^{1}$Institute for Computational Cosmology, Department of Physics, Durham University, South Road, Durham DH1 3LE, UK\\
$^{2}$Centre for Extragalactic Astronomy, Department of Physics, Durham University, South Rd, Durham, DH1 3LE, UK \\
$^{3}$Jet Propulsion Laboratory, California Institute of Technology, 4800 Oak Grove Drive, Pasadena, CA 91109, USA \\
$^{4}$National Astronomical Observatories, Chinese Academy of Sciences, 20A Datun Road, Chaoyang District, Beijing 100012, China \\
$^{5}$School of Astronomy and Space Science, University of Chinese Academy of Sciences, Beijing 100049, China \\
$^{6}$Dipartimento di Fisica e Astronomia “Augusto Righi” - Alma Mater
Studiorum Università di Bologna, via Piero Gobetti 93/2, I-40129
Bologna, Italy \\
}

\date{Accepted XXX. Received YYY; in original form ZZZ}

\pubyear{2022}

\begin{document}
\label{firstpage}
\pagerange{\pageref{firstpage}--\pageref{lastpage}}
\maketitle

\begin{abstract}
 Strong gravitational lensing offers a compelling test of the cold
 dark matter paradigm, as it allows for subhaloes with masses of
 $\sim10^{9}$~M$_\odot$ and below to be detected. We test commonly-used techniques for detecting subhaloes superposed in 
 images of strongly lensed galaxies. For the lens we take a 
 simulated galaxy in a $\sim10^{13}$~M$_\odot$ halo grown in a 
 high-resolution cosmological hydrodynamical simulation, which we 
 view from two different directions. Though the resolution is high, we note the simulated galaxy still has an artificial core which adds additional complexity to the baryon dominated region. To remove particle
  noise, we represent the projected galaxy mass distribution by a
  series of Gaussian profiles which precisely capture the
  features of the projected galaxy. We first model the lens mass as a
  (broken) power-law density profile and then search 
  for small haloes. Of the two projections, one has a regular elliptical shape, while the
  other has distinct deviations from an elliptical shape.  For the
  former, the broken power-law model gives no false positives and correctly
  recovers the mass of the superposed small halo, but for the latter
  we find false positives and the inferred halo mass is
  overestimated by $\sim4-5$ times. We then use a more complex model in which the lens mass is
  decomposed into stellar and dark matter components. In this
  case, we show that we can capture the simulated galaxy's complex projected structures and correctly infer the input small halo.
  
\end{abstract}




\section{Introduction}
The cold dark matter (CDM) model predicts the existence of a vast
population of dark matter haloes, from the scale of galaxy clusters
down to Earth masses and below. Their mass function is
characterised by a simple power law with an exponential cutoff at the
very high mass end \citep{Frenk2012, Wang2020}. For large masses these
predictions have been verified by large
sky surveys \citep{Frenk1990,Rozo2010}. At lower masses, where
dark matter haloes are too small to host a luminous galaxy
\citep{Efstathiou1992, Benson2002, Benitez-Llambay2020}, it remains
unclear whether the prediction still holds true. Alternative dark
matter models predict a cut off of the halo mass function. For
example, warm dark matter (WDM) with a dark matter particle mass of
around a few keV predicts a cut off in the range
$10^6 - 10^{9}\ {\rm M}_\odot$. Pushing constraints on the halo mass
function towards this smaller mass range can distinguish different dark matter models.

Strong gravitational lensing serves as a promising tool to probe the existence of small invisible dark matter haloes. These ``dark'' haloes perturb the images of lensed galaxies when they fall along the path of light from the source to the observer \citep{Koopmans2005, Vegetti2009a, Vegetti2009b}. 
One can statistically study the lensing perturbation of ensembles of small dark haloes, and directly put constraints on the halo mass function \citep{Gilman2019, He2022}. \citet{Gilman2020a} constrained the ``half mode mass''\footnote{This is a characteristic mass related to the $k$-mode where the dark matter power-spectrum has an amplitude half the size expected with CDM. It can be considered as a ``cut-off'' mass, below which the halo mass function is strongly suppressed with respect to CDM.} to be below $10^{7.8}$\,M$_\odot$ by analysing flux ratio anomalies in eight strongly lensed quasar systems. 

Individual subhaloes can be detected by analyzing luminous strong lensing arcs \citep{Vegetti2009b}. \citet{Li2016} has shown that with $\sim 50$ high quality strong lensing images, one can put stringent constraints on the cut-off mass and rule out CDM if no subhalo is detected. More recent work by \citet{Li2017} shows that the existence of line-of-sight haloes can boost the number of detection by a factor of several which improves the constraining power on the identity of dark matter \citep[see also][]{Despali2017, He2021, Amorisco2022}. 

Three such detections of dark haloes with pseudo-Jaffe\footnote{The masses reported are masses of pseudo-Jaffe profiles \citep{Munoz2001}. If modelled by an NFW, the virial mass obtained is usually $0.5\sim1.0$ dex higher than the pseudo-Jaffe mass \citep{Despali2017}.} masses below $10^{10}$~M$_\odot$ have been made so far. The first was made by 
\citet{Vegetti2010}, who detected a $3.51\pm0.15\times10^9$\,M$_\odot$ subhalo using Hubble Space Telescope (HST) imaging. The second was made by \citet{Vegetti2012} via Keck Adaptive Optics, with an inferred mass of $1.9\pm0.1\times10^8$\,M$_\odot$. Finally, \citet{Hezaveh2016} found a $9.1\pm2.5\times10^8$\,M$_\odot$ subhalo using ALMA interferometer observations.
%

%
%
Attempts to detect dark haloes through gravitational imaging have been made in approximately 30 lenses \citep{Vegetti2014, Ritondale2019}, however no other clean detections have been made. Due to the small number of detections, constraints on the halo mass function are somewhat loose, constraining the mass function cut off to be below $\sim 10^{10.9}$~M$_\odot$ \citep{Enzi2021}. This constraint is not yet competitive with other probes, owing to the limited data quality and sample size. Over the next decade, high-quality strong lensing observations from space telescopes such as the James Webb Space Telescope (JWST), Euclid Space Telescope (EST) and China Space Station Telescope (CSST) will allow these constraints to push to much lower cut-off masses on much larger lens samples \citep{Collett2015}.

Substructure detection places stringent requirements on the model of the lens galaxy's mass distribution. The subhalo mass is usually less than $0.1\%$ that of the lens galaxy, necessitating percent level accuracy of the main lens's mass. Inaccuracies in the lens mass model may create ``false-positive'' detections, where the subhalo ``fills-in'' for the mass model's missing complexity. Previous studies have discussed false-positive subhalo detections \citep{Vegetti2010, Ritondale2019}, where they apply strict criteria to ensure all subhalo detections are genuine. This includes requiring a high enough increase of Bayesian evidence that tests on mock data demonstrate the signal cannot be due to an inaccurate mass model \citep{Vegetti2012} and verifying that a consistent subhalo detection is made when pixelised corrections to the gravitational potential are applied \citep{Koopmans2005, Vegetti2009a, Vegetti2010, Vegetti2012}. In certain lenses these potential corrections clearly account for missing complexity in the lens galaxy's mass, thereby correctly flagging a candidate subhalo detection as a false positive. Mass model complexity is not the only contributor to false positives \citep{Vegetti2014}.

This motivates the investigation of more complex lens models, which could improve the subhalo inference by accounting for this missing complexity in the lens galaxy's mass. The subhalo detections listed previously assume a simple parametric model for the lens's mass, the elliptical power-law \citep{Tessore2015} with an external shear (\citet{Hezaveh2016} also included a fourth-order multipole term). However, recent studies have highlighted deficiencies with this model. In a companion paper of this work, \cite{Cao2021} fitted this model to strong lenses simulated using mass models derived from dynamical models of nearby SDSS-IV MaNGA \citep{MaNGA} early type galaxies and showed this can bias the measurement of the local density slopes around the Einstein ring by 13\%. \citet{Gomer2021} and \citet{Van2021} discuss how departures from ellipticity symmetry may affect H$_{0}$ inference in lensed quasars. \citet{Nightingale2019} have also showed that departures from ellipticital symmetry are observed in the luminous emission of three strong lenses.




In this work, we therefore use a hydrodynamic simulation to test the robustness of different parametric lens mass models, focusing on their efficacy for the task of detecting individual subhaloes. By using a simulation, we can compare the lens galaxy's true complex mass distribution to the lens model we fit and if it fails understand why. We perform two tests: (i) we do not add a subhalo to the lens galaxy when generating the mock data and investigate whether a lens model with a subhalo produces a false-positive signal; (ii) we include a subhalo when creating the mock data and test how accurately its mass and position are recovered. We first apply an extension of the commonly used power law profile to fit the main lens \citep{Oriordan2019, Oriordan2020, Oriordan2021}, followed by a ``decomposed'' model which models the lens mass as a combination of stars and dark matter \citep{Dye2005, Nightingale2019}. Our goal is to understand whether modelling the lens mass as a power law profile is sufficient for detecting subhaloes, and if not, whether there is a better model that can provide a correct inference.

Hydrodynamic simulations have previously been used to simulate galaxy-galaxy strong lensing images \citep{Metcalf2014, Xu2017, Mukherjee2018, Despali2020, Enzi2020, Mukherjee2021, Ding2021}. Converting particle data into a corresponding deflection angle field (necessary for lensing) is non-trivial. A common problem is that the mass profiles of galaxies found in hydrodynamic simulations have a sub-kpc core in their centre. The strong lens imaging then produces a bright central image feature, which is not observed in real strong lenses \citep{Bolton2012, Shu2016}. These cores are believed to be due to the limited resolution of the simulations, with previous works assuming a particle resolution of $\sim10^{5}$~M$_\odot$. Our simulation, which has a particle resolution $\sim10$ times that of Illustris-1, still forms a core and central image. We incorporate this feature into our lens modeling such that we can still investigate dark matter subhalo detection. We also mitigate systematic effects related to particle noise in the simulation \citep{Xu2009} and truncation effects which introduce an artificial shear \citep{Van2020, Ding2021}.

Testing with galaxies from the Illustris simulation \citep{Vogelsberger2014}, \citet{Xu2017} demonstrated that deviations of simulated galaxies from a simple elliptical power-law profile affect inference on the Hubble constant.  More recently, \citet{Enzi2020} used 10 galaxies from the Illustris-1 simulation to test the power-law lens assumption for substructure lensing, and showed no degeneracy between the complexity of the true mass distribution of their mock lenses and the inferred substructure abundance. However, their work focused on the statistical properties of subhaloes' signals and did not test individual subhalo detection. To fully understand how the use of simple parametric lens models affects the detection of individual substructure, testing with mock lenses extracted from simulations is necessary.

This paper is structured as follows: in Section~\ref{sec:mock}, we introduce our simulation data and the way we simulate strong lensing images from particle data. In Section~\ref{sec:pl_model}, we introduce how we model the lensing images and search for subhaloes. In Section~\ref{sec:pl_results}, we show the power law fitting results. In Section~\ref{sec:star_dark}, we introduce a more complex lens model where we model the lens' stellar and dark components separately and then we show how it behaves for our tests. In Section~\ref{sec:discussion}, we discuss our results. Finally, in Section~\ref{sec:conclusions}, we summarize our results. All the computations, if not specified, are done by the state-of-art open source strong lensing software \textsc{PyAutoLens}\footnote{The {\tt PyAutoLens} software is open-source and available from \url{https://github.com/Jammy2211/PyAutoLens}.} \citep{Nightingale2018, pyautolens}. Throughout the paper we adopt the Planck cosmology \citep{Planck2016}, of which $\rm H_0 = 67.7 \; km\,s^{-1}\,Mpc^{-1}$, $\rm \Omega_m = 0.307$ and
$\rm \Omega_\Lambda = 0.693$. 

\section{Mock Lensing Images}\label{sec:mock}


\subsection{Particle data}

We create our lens galaxy by using data from a cosmological hydrodynamical zoom-in simulation of a $\sim10^{13}$\,M$_\odot$ galaxy group \citep{Richings2021}. The simulated galaxy is selected from the EAGLE 100 Mpc box-size simulation \citep{Schaye2015} and was first identified by \citet{Despali2017} as having similar properties to lenses from the Sloan Lens ACS (SLACS) survey \citep{Bolton2006}. The friends-of-friends (FOF) ID of the halo is 129. To resolve dark matter haloes with masses down to $\sim10^6$\,M$_\odot$, this zoom-in simulation applies a novel technique whereby there are many more dark matter particles than gas particles. Unlike the common construction of initial conditions in hydrodynamic simulations, where each dark matter particle in a dark-matter-only simulation is split into a pair of dark matter and gas particles, the simulation we use initializes 7 dark matter particles per gas particle, resulting in dark matter and gas particle mass of 8.3$\times$10$^4$\,M$_\odot$ and 10.7$\times$10$^4$\,M$_\odot$, respectively. At $z=0$, the dark matter halo of the zoom-in simulation's galaxy group has a mass of $m_{200}=10^{13.14}$~M$_\odot$ and size of $r_{200}=506$~kpc.\footnote{$m_{200}$ is the mass enclosed within a radius of $r_{200}$, where $r_{200}$ is determined as the radius at which the mean enclosed density is 200 times the critical density of the universe.} The Plummer-equivalent gravitational softening length is 0.05~kpc.

\begin{figure}
	\includegraphics[width=1.0\columnwidth]{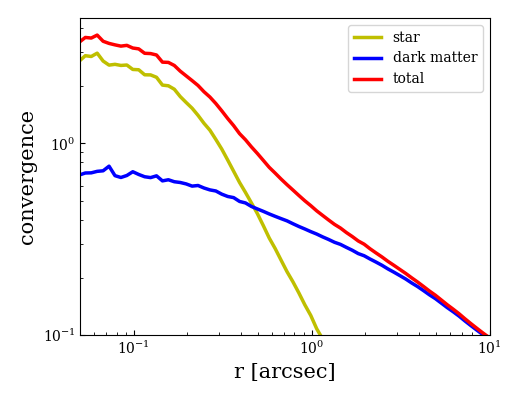}
    \caption{Convergence of different components of the simulated galaxy. Yellow, blue and red represents the stars, dark matter and total matter respectively, assuming the lens at z = 0.2 and source at z = 2.5. At the redshit of the lens $z = 0.2$, 1~arcsec corresponds to 3.3~kpc in angular size.}
    \label{fig:profiles_3D}
\end{figure}

In Fig.~\ref{fig:profiles_3D}, we show the convergence profile (projected density divided by a constant lensing critical density) of different components of the simulated galaxy assuming the lens and source galaxy to be at $z=0.2$ and 2.5, respectively. Inside the central $~\sim0.7$~arcsec, the baryonic mass is larger than the dark matter mass, and the central density of stellar mass is around 4 times higher than that of the dark matter. A constant-density core with a size of $\sim0.1$ arcsec exists in the central region, which is a result of the finite resolution of the simulations. This phenomenon has been seen in several other studies that simulate strong lens images from simulation data \citep{Mukherjee2018, Enzi2020, Ding2021}, and it can produce a dim central image in simulated strong lensing images that is rarely seen in real observations \citep{Winn2004, Quinn2016}. The core feature introduces additional complexity beyond realistic massive ellipticals and thus might lead to an overestimation on the baryonic effects in our tests. Fortunately, for the lens configurations considered in this work, the central image is sufficiently dim and small that one can mask it out without it impacting the lens modelling and subhalo inference, an approach also followed by \citet{Enzi2020}. To do this, we artificially increase the assumed error on the flux in the region containing the central image to such high values that they are effectively removed from the goodness-of-fit measurement.\footnote{We previously attempted to remove the central image by applying a mask that removed the data in the central region altogether. However, we found that this introduced systematics due to edge effects associated with the source-plane pixelisation. Pixels at the edge of the mask (which have non-negligible flux due to the central image) were not appropriately regularized because their neighbours were not traced to the source plane \citep[see][]{Nightingale2018}.} 

\subsection{Simulating strong lensing images}

\subsubsection{Mock lenses}
To simulate images that are strongly lensed by the particle distribution from a hydrodynamical simulation one needs a method which can determine the corresponding deflection angle map. There are two common ways of approaching this: (i) derive the projected density distribution of the particle data and solve for its potential via a Fast Fourier Transform (FFT) or; (ii) assume analytic profiles representing each particle enabling deflection angles to be easily computed, such that the overall deflection field is the sum over all particles. For the latter method, the computational cost can be greatly reduced by using a $k$-d tree algorithm \citep{Bentley1975}, making it faster than the FFT method at comparable resolution \citep{Metcalf2014,Pekova2014}. However, neither method offers a well posed way of quantifying particle noise in the deflection angles, which can closely resemble the deflection angles of a dark metter subhalo in a strong lens \citep{Xu2009}. Besides the particle noise, these above methods also face the boundary truncation effect, which is that when truncating the particle data in an improper way (e.g. a square boundary applied to an elliptically shaped galaxy), an artificial shear component is introduced \citep{Van2020}. The shear magnitude depends on the galaxy's profile, the truncation area size and the truncation scheme used, and an improper truncation on the particle mass data can induce several percent bias to H$_0$ inference \citep{Van2020, Ding2021}. 

To avoid particle noise and the boundary truncation effect, we therefore instead fit analytic profiles to the simulated galaxy's particle data and use these profiles to compute our lens galaxy's deflection angle field. We approximate the projected mass distribution of the simulated galaxy using the multiple Gaussian expansion (MGE) method, which is widely used for modelling galaxy surface brightness profiles in studies of galaxy stellar dynamics \citep[e.g.][]{Cappellari2008, Hongyu2016, Li2019, He2020}. \citet{Hongyu2016} applied the expansion method to both galaxies and dark matter haloes in Illustris simulation, showing that it has flexibility to capture irregular and asymmetric features in a galaxy's light or mass distribution. The deflection angles (and other lensing quantities) of an elliptical Gaussian profile can be easily computed \citep{Shajib2019}, making lens simulations convenient and fast.


\begin{figure}
	\includegraphics[width=1.0\columnwidth]{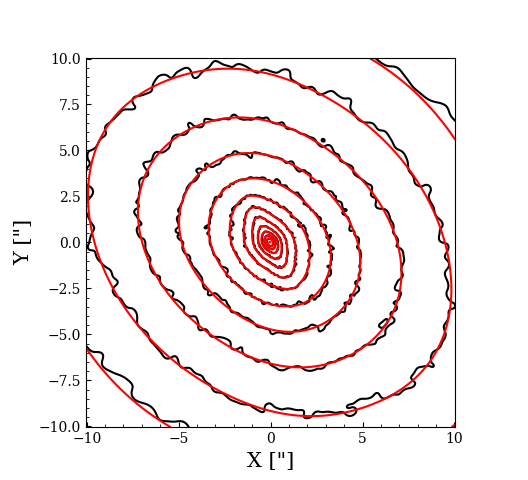}
    \caption{Iso-density contours of the projected stellar mass distribution of our simulated galaxy. The contours measured directly from the particle data using GLAMER \citep{Metcalf2014, Pekova2014} are shown in black, while those for the best-fit MGE are in red. The contours are evenly log-spaced in projected density. From inside to outside, each contour decreases by 0.4 order. Note that only very inner part (within 5 arcsec) of this image is observable and the reason we plot it on a much larger region is to show that the MGEs represent the lens' stellar mass well to a very large range. For clarity, later plots are all zoomed in to the region around the Einstein radius.}
    \label{fig:mge_represent}
\end{figure}

We compute the deflection angles of the simulated galaxy separately for its stellar and dark matter components, and then add them together to get the total deflection angles. The gas component is omitted because its contribution to the total mass in the galaxy's central region is negligible. We will add subhaloes to the deflection angle map via an analytic mass profile and therefore must ensure no subhaloes in the particle data are included in our simulation process. We therefore only use particles belonging to the main halo identified by the SUBFIND algorithm \citep{Springel2001}.

We set the lens galaxy to be at redshift $z$ = 0.2, and the source galaxy at $z$ = 2.5. We first use \textsc{GLAMER} \citep{Metcalf2014, Pekova2014} to generate convergence maps, where each particle is represented by a smoothed b-spline in 3D. For each star particle the smoothing length is the distance to its 8th nearest (stellar) neighbour and for each dark matter particles it is the distance to its 64th nearest (dark matter) neighbour. We then use the MGE code of \citet{Cappellari2002} to decompose the convergence maps into multiple Gaussian profiles, where the Gaussian components share the same centre but are free to have different amplitudes, sizes, position angles and axis ratios. As an example, Fig.~\ref{fig:mge_represent} shows contours tracing the particle data input (dark lines) and MGE best-fit (red lines) of the simulation's projected stellar mass distribution. The MGE-fitting code decomposes the stellar component into 13 individual Gaussian profiles and Fig.~\ref{fig:mge_represent} shows asymmetric features such as the twist in ellipticity are well captured by the MGEs. The relative errors between the input profile and best-fit MGEs are smaller than $\sim5\%$. We apply the same routine to the simulation's dark matter particles and then add the best-fit Gaussian profiles together to represent the simulated galaxy's total projected mass distribution.

\begin{figure*}
	\includegraphics[width=2.0\columnwidth]{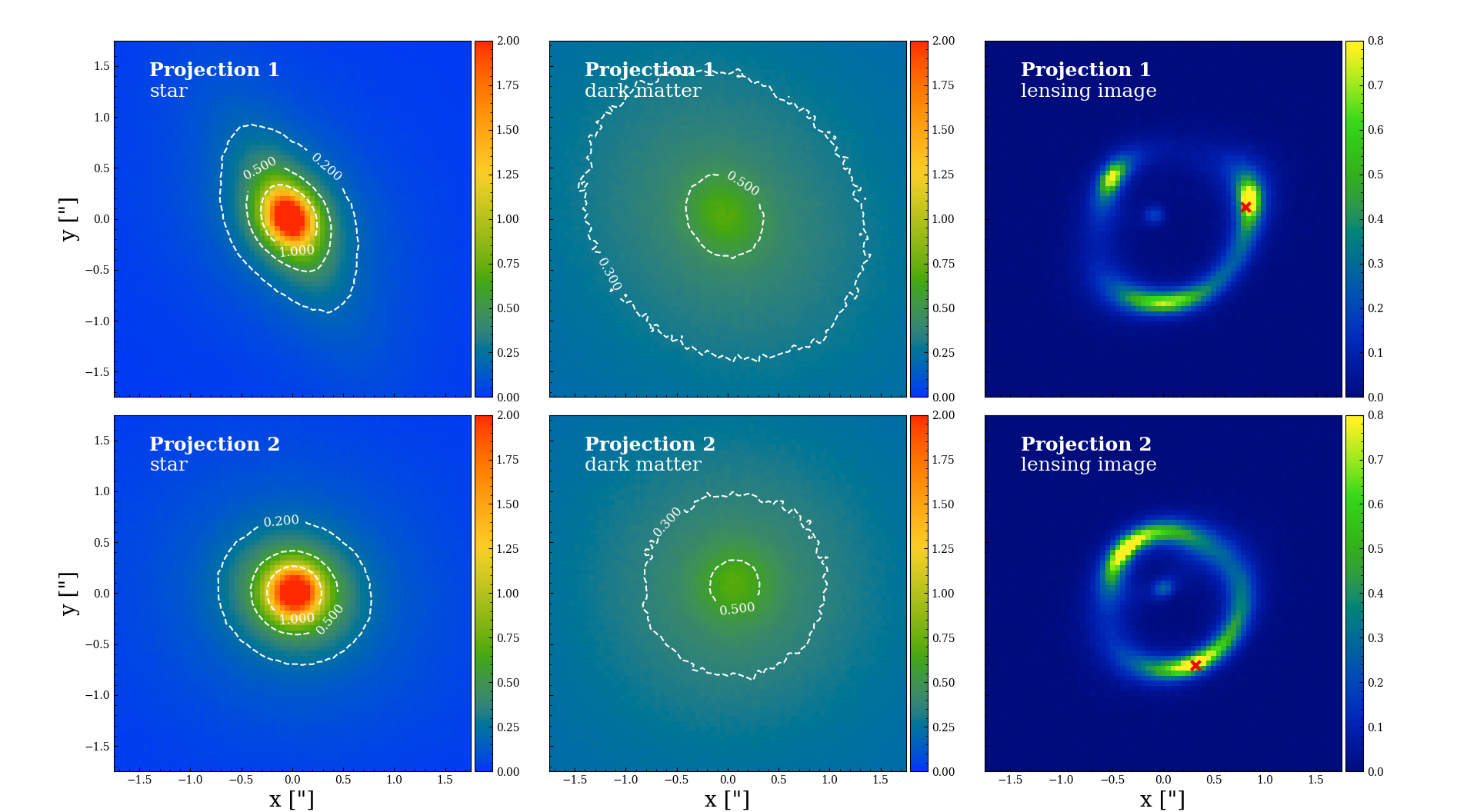}
    \caption{Input stellar (left column) and dark matter (middle column) convergence maps of our simulated galaxy along two different lines of sight (the two rows).  The corresponding strong lensing images are shown in the right column. The top row is for the line of sight that produces pointy shaped iso-convergence contours, and the bottom row shows the projection with a rounder convergence field. The red crosses on the mock images mark the positions where we will later place subhaloes.}
    \label{fig:cvgc_mocks}
\end{figure*}

In the top-left panel of Fig.~\ref{fig:cvgc_mocks}, we show the convergence (i.e. the projected density divided by the critical surface density for lensing) of the MGE representation of the simulated galaxy, where a pointy ``American football-like'' shape can be seen. To investigate how lens model fits change depending on the shape of the convergence, we rotate the same galaxy to view it along a different line of sight, intentionally choosing a viewing angle that produces a rounder convergence map, which is shown on the bottom left panel of Fig.~\ref{fig:cvgc_mocks}. Following equation~(43) of \citet{Shajib2019}, we compute MGE deflection angles for both projections, which are then used to simulate strong lensing images. For an accurate computation of each image pixel's flux, we treat every pixel with a $4\times4$ subgrid so that for each image pixel 16 light rays are traced to the source plane, with the pixel flux set to their mean value.

\subsubsection{Mock sources}

We simulate source galaxies using the cored S\'ersic profile \citep{Trujillo2004},
\begin{equation}\label{equ:core_sersic}
    I(r) = I^{'}{\rm exp}\left[- b_n \left(\frac{r^2 + r^2_{\rm c}}{r_{\rm e}^2}\right)^{1/(2n)}\right],
\end{equation}
where $I^{'}$ is the scale density, $r_{\rm c}$ is the core size, $r_{\rm e}$ is the effective radius, $n$ is the S\'ersic index and $b_{n}$ is a dimensionless parameter fully determined by $n$ \citep{Graham2005}. Our input sources are a single elliptical cored S\'ersic profile (the ellipticity is introduced using $r=\sqrt{\left(x/q\right)^2 + y^2}$), which is simple compared to observed lenses where source galaxies are more complex and show features such as multiple star forming clumps, spiral structures and extremely compact centres. Our choice to assume a simple source profile is to make it straight forward to test the effects of using different lens mass models. Our lens modelling procedure uses pixelised source reconstructions which are able to fit the more complex sources seen in real data \citep{Nightingale2019}.


\subsubsection{Subhaloes}
Some of our mock lensed images include a dark matter subhalo in the lens galaxy near one of the arcs. We represent subhaloes using the spherical NFW profile \citep{NFW1996}
\begin{equation}\label{equ:NFW}
    \rho(r) = \frac{M_0}{4{\rm \pi}r\left(r_{\rm s} + r\right)^2},
\end{equation}
where $M_0$ is the scale mass and $r_{\rm s}$ is the scale radius. Following equation (A.18) of \citet{Baltz2009}, we analytically compute its deflection angles and add it to that inferred via the MGE fit to the stellar and dark components. We assume the NFW halo follows the mass-concentration relation given by \citet{Ludlow2016}, allowing us to parameterize it with only its mass, $m_{200}$ (and its position). Mock lenses are generated with an input subhalo of two different masses, $m_{200}=5\times10^8$~M$_\odot$ or $m_{200}=5\times10^9$~M$_\odot$. 

\subsubsection{Data Quality}

We simulate mock images similar to observations of the Hubble Space Telescope/Advanced Camera for Surveys (HST/ACS) Wide Field Camera, with a pixel size of 0.05\arcsec and a Gaussian point spread function (PSF) with a standard deviation of 0.05\arcsec \footnote{The actual PSF $\sigma$ for HST/ACS is $\sim$0.034\arcsec and the pixel size is 0.04\arcsec, so our tests are slightly worse than real HST observations in terms of resolution. However, in terms of the PSF modelling, we assume we have perfect knowledge of the PSF, which goes in the other direction of being optimistic.}.
We set the normalisation of the source's surface brightness to give a signal-to-noise ratio (S/N) of $\sim80$ in the brightest pixel of the lensed source's image, whilst adjusting the background noise level to closely match that expected from a few HST orbits. For our mocks, the background sky noise is 0.1 ${\rm e}^{\rm -}{\rm pixel}^{\rm -1}{\rm s}^{\rm -1}$. This S/N represents observations that are around double the highest S/N sources observed currently with Hubble, for example the SLACS sample. Using such high S/N data is a choice we made to ensure our tests of deficiencies in the lens mass model are easier to distinguish from noise in the mock data. The right column of Fig.~\ref{fig:cvgc_mocks} shows the two mock images, where a source galaxy is lensed by the two different line-of-sight projections (the corresponding projected densities are shown on the left). For mock datasets which include a subhalo, the positions marked by red crosses showing the locations of the subhalo that we add. In Table~\ref{tab:mock_images}, we summarize the relevant parameters used to simulate these images.

\begin{table}
    \renewcommand\arraystretch{1.5}
	\centering
	\begin{tabular}{ccc} 
	\hline
	& \textbf{Projection 1} & \textbf{Projection 2} \\
	\hline
	\textbf{Input Lenses} & & \\
	\hline
    Stellar MGE number & 13 & 11 \\
	Dark MGE number & 5 & 5 \\
	redshift & \multicolumn{2}{c}{0.2} \\
	\hline
	\textbf{Input Sources} & \multicolumn{2}{c}{\textbf{Cored S\'ersic}} \\
	\hline
	centre(x, y) [(\arcsec, \arcsec)] & \multicolumn{2}{c}{(0.08, -0.03)} \\
	axis ratio & \multicolumn{2}{c}{0.55} \\
	position angle [$^\circ$] & \multicolumn{2}{c}{30} \\
	$I^{'}$ [e$^{-}$ pix$^{-1}$ s$^{-1}$] & \multicolumn{2}{c}{2.0} \\
	$r_{\rm e}$ [\arcsec] & \multicolumn{2}{c}{0.11} \\
	$n$ & \multicolumn{2}{c}{2.0} \\
	$r_{\rm c}$ [\arcsec] & \multicolumn{2}{c}{0.01} \\
	redshift & \multicolumn{2}{c}{2.5} \\
	\hline
	\textbf{Input Subhaloes (if added)} & \multicolumn{2}{c}{\textbf{Spherical NFW profile}} \\
	\hline
	centre(x, y) [(\arcsec, \arcsec)] & (0.81, 0.12) & (0.32, -0.71) \\
	$m_{200}$ [M$_\odot$] & \multicolumn{2}{c}{$5\times10^8$ or $5\times10^9$} \\
	mass-concentration relation & \multicolumn{2}{c}{\citet{Ludlow2016}} \\
	redshift & \multicolumn{2}{c}{0.2} \\
	\hline
	\textbf{Image Settings} & & \\
	\hline
	pixel size [\arcsec] & \multicolumn{2}{c}{0.05} \\
	PSF $\sigma$ [\arcsec] & \multicolumn{2}{c}{0.05} \\
	background noise level [e$^{-}$ pix$^{-1}$ s$^{-1}$] & \multicolumn{2}{c}{0.1} \\
	exposure time [s] & \multicolumn{2}{c}{8000} \\
	maximum pixel S/N & \multicolumn{2}{c}{$\sim$ 80} \\
	\hline
	\end{tabular}
	\caption{Parameters used to simulate the mock lensing images.}\label{tab:mock_images} 
\end{table}

\section{Method}\label{sec:pl_model}

\subsection{Mass Models}
\subsubsection{Broken Power Law}

The simulated lens galaxy has an artificial $\sim0.1$\,kpc constant density core, which forms a spurious central, demagnified image. We mask this central image by manually decreasing the contribution of central image pixels to the likelihood calculation, but must also ensure our mass model parameterization is able to represent the cored density, to avoid biasing our reconstruction of the lensed source's arcs \citep{Enzi2020}. We therefore assume the elliptical broken power law (eBPL) profile \citep{Oriordan2019, Oriordan2020, Oriordan2021} with convergence 
\begin{equation}\label{equ:power_law}
    \kappa{\left(r\right)}=\left\{
    \begin{array}{ll}
    \kappa_{\rm b}\left(r_{\rm b} / r \right)^{t_1}, & r \leq r_{\rm b} \\
    \kappa_{\rm b}\left(r_{\rm b} / r \right)^{t_2}, & r > r_{\rm b}
    \end{array}
    \right.,
\end{equation}
where $r_{\rm b}$ is the break radius, $\kappa_{\rm b}$ is the convergence at the break radius, $t_1$ is the inner slope and $t_2$ is the outside slope. When $r_{\rm b}=0$, the eBPL reduces to the standard power law profile with 3D density $\rho(r)\propto r^{-\gamma}$, as used in many lens studies. We introduce ellipticity by setting $r = \sqrt{q x^2 + y^2 / q}$, where $q$ is the axis ratio. In practice, we parameterize a model's axis ratio and position angle, $\theta$, in terms of two components of ellipticity:
\begin{align}
e_{1} &=\frac{1-q}{1+q} \sin 2\theta~, &
e_{2} &=\frac{1-q}{1+q} \cos 2\theta~.    
\end{align}
With an additional two parameters describing the profile's centre, the eBPL model has 8 free parameters. Degeneracies between certain parameters in the eBPL profile, for example the two different slopes, make it challenging to fit efficiently and avoid inferring local maxima. We therefore assume priors that lessen these degeneracies and simplify parameter space, where we constrain $r_{\rm b} \leq 0.4\arcsec$, $t_{1} \leq 0.5$ and $t_2 > 0.5$. For some cases, we further limit the Einstein radius to be larger than 0.5\arcsec. All eBPL models are fitted with an additional external shear in the lens model, which provides further flexibility in stretching and squeezing of the mass profile that can capture asymmetric features in the lens's convergence \citep{Cao2021}. Similar to the ellipticity parameterization, the external shear is also parameterized with two components $\gamma_{\rm 1ext}$ and $\gamma_{\rm 2ext}$, where the shear's magnitude, $\gamma_{\rm ext}$, and position angle, $\theta_{\rm ext}$, can be recovered as
\begin{align}
    \gamma_{\text{ext}} &= \sqrt{\gamma_{\text{1ext}}^2+\gamma_{\text{2ext}}^2}~, &
    \tan{2\theta_{\text{ext}}}&=\frac{\gamma_{\text{2ext}}}{\gamma_{\text{1ext}}}~.
\end{align}
For modelling of a subhalo, we take the same NFW form we use to simulate the image.

\subsubsection{Source reconstruction}

The final lens models of our analysis -- from which all results in the main content of this paper are taken -- reconstructs the source galaxy using a pixelisation that adapts to the source's surface brightness distribution \citep[see][for a discussion of systematics this approach removes compared to other pixelisations]{Nightingale2018}. However, before using this pixelised source, a number of initial fits are performed which estimate the parameters of the lens mass model efficiently (for details see the next subsection). These fits assume either a parametric source which is modelled using the S\'ersic profile ($r_{\rm c} = 0$ in Eq.~\ref{equ:core_sersic}) or a pixelised source where the density of pixels adapts to the magnification, leading to smaller pixels in more magnified areas of the source plane. In appendix \ref{AppendixA} we also show that we reproduce our main conclusions assuming a parametric cored-S\'ersic source model.

\subsection{Fitting procedure}

We use {\tt PyAutoLens} \citep{pyautolens} to model the simulated lens datasets, which is described in \citet[][N18 hereafter]{Nightingale2018} and builds on the works of \cite{Warren2003, Suyu2006, Nightingale2015}. {\tt PyAutoLens} uses a technique called ``non-linear search chaining'' to compose pipelines which break the lens modelling procedure into a series of simpler model fits. This allows us to begin modelling our data with a simple lens model (e.g. an isothermal mass profile and a S\'ersic source) and via a sequence of non-linear searches gradually increase the model complexity, so as to eventually fit the desired more complex lens model (in this work, mass models which include a dark matter subhalo and with a source reconstructed on the brightness-based pixelisation). Non-linear search chaining is implemented in {\tt PyAutoLens} via the probabilistic programming language {\tt PyAutoFit}\footnote{\url{https://github.com/rhayes777/PyAutoFit}} \citep{pyautofit}. We use the nested sampling algorithm \texttt{dynesty} to perform all model fits.

\begin{figure}
	\includegraphics[width=1.0\columnwidth]{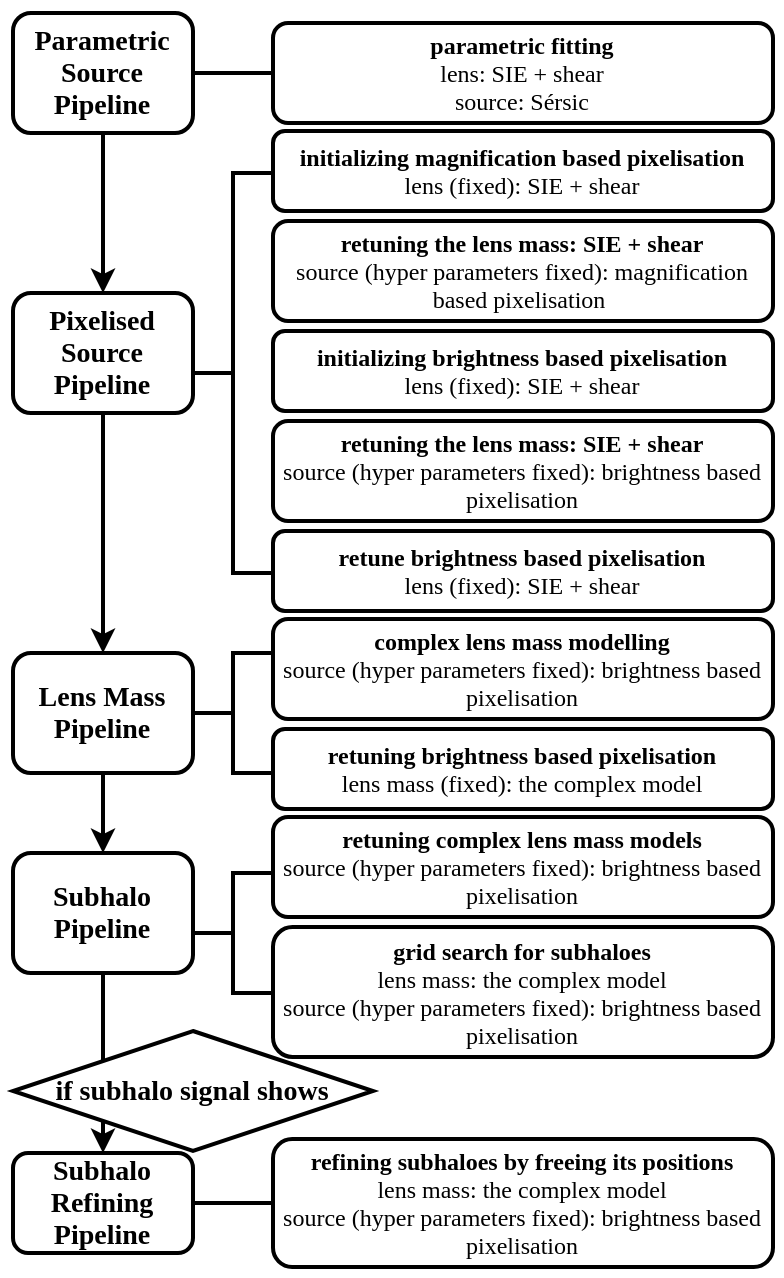}
    \caption{The fitting procedure we use to search for subhaloes.}
    \label{fig:lensing_procedure}
\end{figure}

We employ the Source, Light and Mass (SLaM) pipelines that are distributed with {\tt PyAutoLens} \footnote{\url{https://github.com/Jammy2211/autolens_workspace}}. The SLaM pipelines were used in the work of \citet{Cao2021} and \citet{Etherington2022} and our analysis closely follows theirs, albeit we end with an additional pipeline that determines whether including a subhalo in the lens model increases the Bayesian evidence relative to the model without a subhalo. Like in \citet{Cao2021}, we do not need to model the lens light and therefore employ a model fitting procedure consisting of four distinct pipelines which each focuses on fitting a specific aspect of the model. These pipelines are, in order: (i) the parametric source pipeline; (ii) the pixelised source pipeline; (iii) the lens mass pipeline and; (iv) the subhalo pipeline. Each pipeline consists of one or more non-linear searches that fit a unique lens model parameterization, which  Fig.~\ref{fig:lensing_procedure} shows a flow chart of, which we will now explain in detail.

\subsubsection{Parametric source pipeline}
The parametric source pipeline aims to initialize a robust model for the lens galaxy's mass by fitting a source galaxy that has a smooth analytic form. The primary reason for this pipeline is that a robust model for the lens galaxy's mass is necessary to avoid pixelised source reconstructions inferring the unphysical solutions described by \citet{Maresca2021}, where the reconstruction inferred is a demagnified version of the lens data. This pipeline assumes a singluar isothermal ellipsoid (SIE) mass model (where in Eq.~(\ref{equ:power_law}) $r_{\rm b}$ is set to be 0 and $t_2$ is fixed to be 1.0) with an external shear and a S\'ersic profile for the source surface brightness.

\subsubsection{Pixelised source pipeline}
The pixelised source pipeline is composed of four search phases. The first search fits for parameters describing the resolution of the magnification based pixelisation and the regularization coefficient of the constant regularization scheme, with the lens mass model fixed to the result of the parametric source pipeline. The second search re-fits the lens mass model using the pixelisation and regularization inferred previously. The third search fits for parameters that derive the surface brightness based pixelisation and the luminosity weighted regularization scheme, where the  lens mass model is fixed to the best-fit values inferred in the previous search. The fourth search again re-optimizes the lens mass model now using the brightness based pixelisation and regularization and we finally re-fit the pixelisation and regularization parameters again one last time, ensuring that the source reconstruction is tailored to the properties of the source it is fitting.

\subsubsection{Mass pipeline}
This pipeline fits a more complex lens mass, either the eBPL model plus an external shear or the decomposed model that separately models the stellar and dark components plus an external shear. This pipeline consists of two searches. It first fits the new lens mass model with fixed source pixelisation parameters. The priors of the (broken) power law model's centres, elliptical components, Einstein radius are updated using information of the previous best-fit models. We set those priors to be Gaussian priors centering on corresponding best-fit values of previous models and their widths are set manually using values which balance reducing the size of parameter space to ensure an efficient fit whilst being broad enough not to remove physically plausible solutions. For other parameters (like the break radius, inner (outer) slopes, and external shear) we assume broad uniform priors that are not informed by the previous mass model fits. Having now fitted this more complex mass model, we again update the source pixelisation and regularization parameters using the best-fit lens mass model of the first step. This is the final fit which updates the pixelisation and regularization parameters, with all remaining fits focusing on the lens (and subhalo) mass models.

\subsubsection{Subhalo pipeline}\label{sec:SubhaloPipeline}

This pipeline performs Bayesian model comparison to determine if a lens model with a subhalo is preferred over a lens model without a subhalo. The pipeline begins by fitting the same lens mass model (with fixed source pixelisation and regularization parameters) inferred at the end of the mass pipeline, with all priors inherited from this fit. This provides us with an estimate of the Bayesian evidence of the lens model without a subhalo. We then fit lens models which include an NFW subhalo. For the subhalo's mass, we assume a uniform prior on $\log_{10}\left(m_{200} / {\rm M}_{\odot}\right)$ between 6 and 11. 

Due to the complexity of our parameter space (which consists of the mass models of both the main lens galaxy and a subhalo) we found it was common for the inferred posterior to correspond to a local likelihood maximum (as opposed to the global maximum). To mitigate this, we scan for subhaloes using a grid of non-linear searches, where each search confines the $(x,y)$ image-plane coordinates of the subhalo to a small 2D square segment of the image-plane. We perform $25$ independent model fits, corresponding to a $5 \times 5$ grid, which divides the image region between $-1.0\arcsec$ and $1.0\arcsec$ into sub regions with sizes of $0.4\arcsec\times0.4\arcsec$. The parameters of the main lens are fit for simultaneously along with the subhalo parameters in each of these 25 fits.

\begin{figure*}
	\includegraphics[width=2.0\columnwidth]{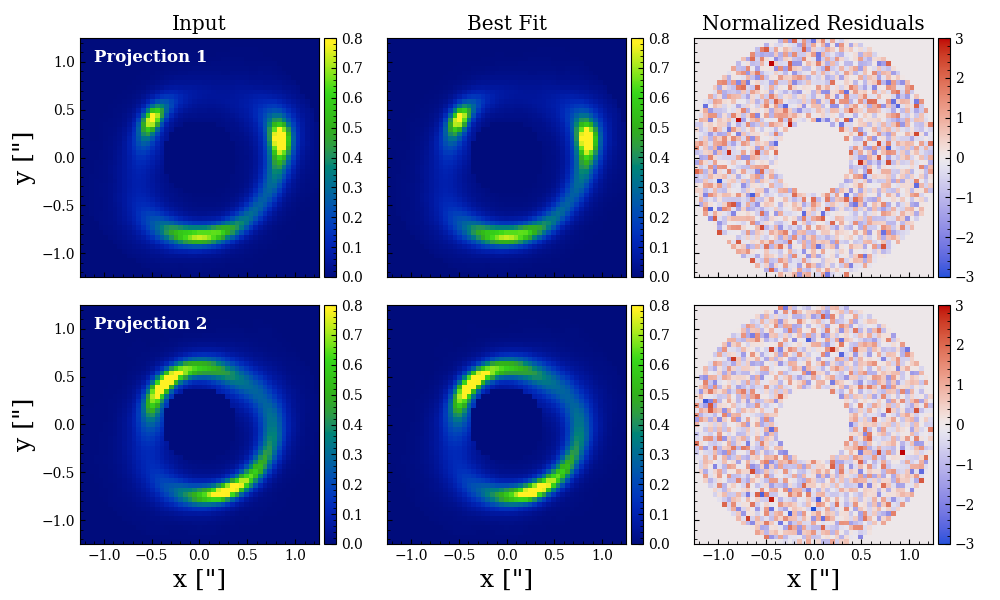}
    \caption{\textbf{Left column:} the mock lensing images. \textbf{Middle column:} the best recovered images using an eBPL to model the lens mass. \textbf{Right column:} the corresponding normalized residuals (residuals divided by the noises). The top row show the case of Projection 1 and the bottom row show the case of Projection 2. The color bar unit for the left two column images is e$^{-}$~pix$^{-1}$~s$^{-1}$. The units of the y and x axes are arcsec.}
    \label{fig:fit_images}
\end{figure*}

To determine whether the lens model with a subhalo is favoured by the data over the model without a subhalo, we must choose a statistical quantity with which to compare them. Obvious choices are the Bayesian evidence or differences in maximum log likelihood values. We use the maximum log likelihood to compare models which do and do not include a subhalo. However, the Bayesian evidence is as an output of \texttt{dynesty} and we have verified that our results are unchanged using this quantity.
We denote the difference between the two maximum log likelihoods to be $\Delta L$, such that if $\Delta L$ in certain cells of the subhalo search are large it suggests the existence of a subhalo within one of those certain grids. Instead, if all log likelihood differences are very small, then it indicates no subhaloes of a sufficiently high mass to be detected are present in the image. For this paper, we take the threshold as follows: if $\Delta L \leq 5$, we call it a non-detection; if $5 < \Delta L \leq 10$, we call it a plausible detection; if $\Delta L > 10$, we call it a detection.  

If the subhalo grid search has a plausible detection ($\Delta L > 5$) the subhalo pipeline performs one more fit, which fits for both the main lens and subhalo parameters. The subhalo's $(x,y)$ position is no longer confined to a square segment of the grid search and we instead place a Gaussian prior on the $x$ and $y$ positions. The 2D Gaussian prior is centred at the maximum-likelihood subhalo position inferred previously using the grid search, with a relatively large standard deviation of 0.5\arcsec. For the subhalo's mass, we retain a prior uniform in log$_{10}M$ between $10^6$~M$_\odot$ and $10^{11}$~M$_\odot$.

\section{Power Law Tests}\label{sec:pl_results}

We first use our simulated lenses to test the broken power law profile, which is commonly assumed in strong lensing studies to model the mass distribution of the lens galaxy \citep{Vegetti2014, Collett2014, Dye2015, Ene2018}. Our tests are divided into two parts: (i) how do power-law fits behave for the case where no subhalo is present in the mock data; (ii) can the power-law correctly recover the subhalo's properties when there is one present in the mock data. For convenience, we call the tests where no subhalo is added ``smooth tests'', and tests where there is a subhalo added  ``subhalo tests''.     

\begin{figure*}
	\includegraphics[width=2.0\columnwidth]{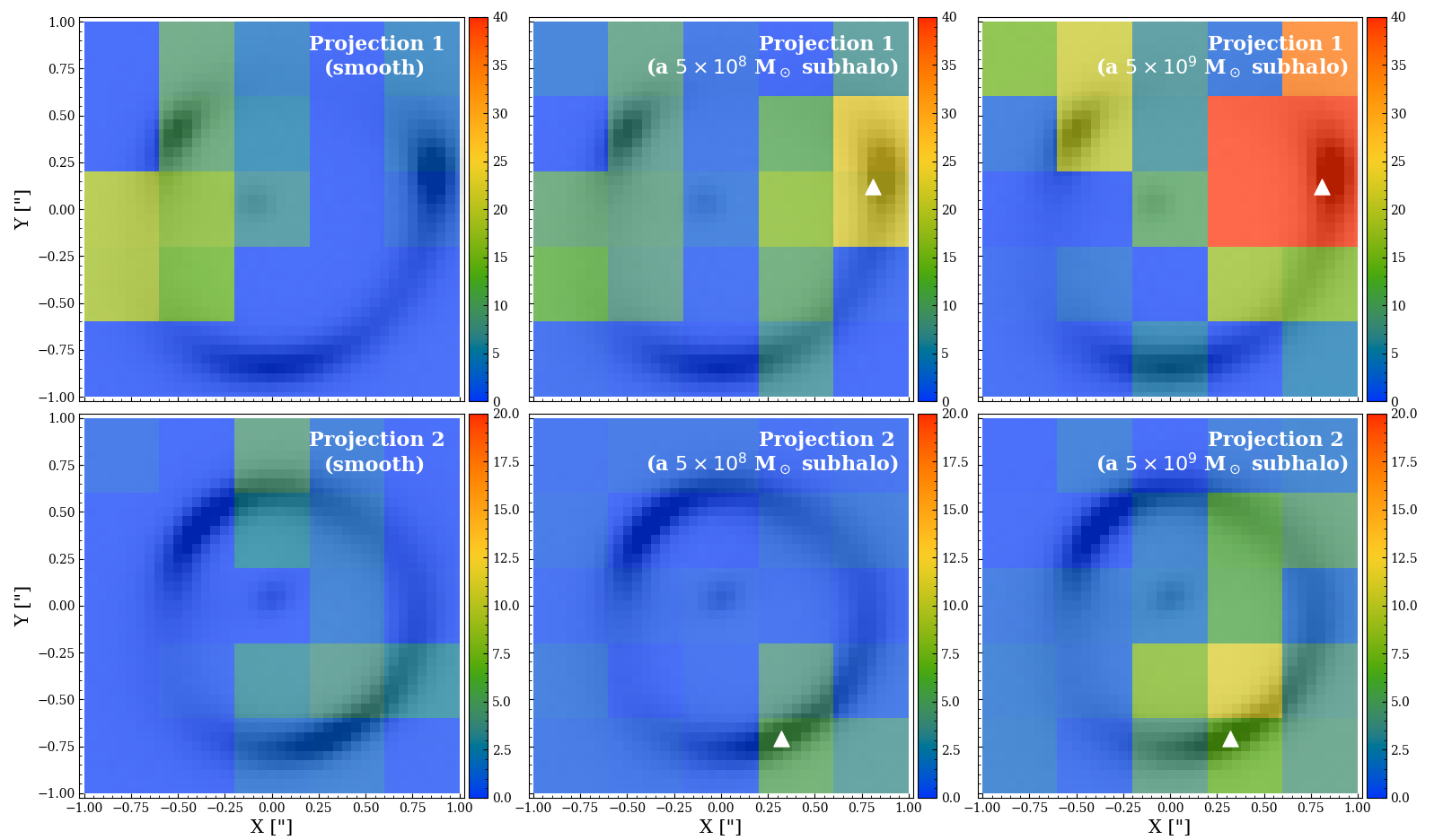}
    \caption{When modelling the lens mass distribution with an eBPL, substructures are (too) easily detected. Colours indicate the increase in maximum log likelihood, $\Delta L$, when a subhalo is included inside $0.4\arcsec\times0.4\arcsec$ squares during a fit to a lens that has: no subhaloes (left), a subhalo of mass $5\times10^8$~M$_\odot$ (middle), or a subhalo of mass $5\times10^9$~M$_\odot$ (right). Top and bottom rows show the results for Projection 1 and 2 (with different colour scales). White triangles mark the true locations of the subhaloes. Note the false-positive detections in the left panels; the best-fit subhalo masses in the other panels are also overestimated by a factor four to five.}
    \label{fig:pl_smth_lklhd}
\end{figure*}

\subsection{Smooth test results}

In Fig.~\ref{fig:fit_images}, we show the input and recovered images from the best-fit smooth model when fitted to simulated images that do not include a subhalo (this corresponds to the first model fit in the subhalo pipeline and will act as the model we compare to models including a subhalo in a moment). For visual clarity, we have removed the central image caused by the core of the simulated galaxy, however note that this region is included in the model-fit with high error values. For both projections, the reconstructed images in the middle panel are similar to the input images shown on the left. The normalized residuals (residuals divided by the noises) shown in the right panel confirm the good fit, showing no clear or obvious correlated residuals. It is noted that the best-fit eBPL model's break radius for projection 1 and 2 are $\sim0.2$\arcsec and 0.1\arcsec respectively, which confirms that the core is able to affect the lensing even though the central image has been masked out. Using an eBPL model is therefore necessary to account for the core.

We now consider the results of the subhalo search. The left column of Fig.~\ref{fig:pl_smth_lklhd} shows the results of the subhalo phase, using the quantity $\Delta L$ (defined in Section~\ref{sec:SubhaloPipeline}) inferred in every cell of the subhalo-position grid. The upper and lower panels show the results of Projection 1 and 2 respectively. For Projection 1, where the input galaxy has a pointy shaped convergence, grids around the top-left luminous arc have $\Delta L$ over 10, and the highest $\Delta L$ is $\sim$ 21.4 for the left most grid cell of the third row from bottom. For the grid cell with $\Delta L\sim$ 21.4, a subhalo with $m_{\rm 200}$ of $10^{9.8^{+0.4}_{-0.5}}$~M$_\odot$ is inferred around that region. Given that the simulated lens galaxy we fitted here does not contain a subhalo, this signal is a false-positive. However, for Projection 2 which has a rounder convergence, no grid has a $\Delta L > 5$. Assuming our criteria of requiring $\Delta L > 5$ the inclusion of an additional subhalo model using the eBPL is therefore correctly not favoured by the data and the eBPL gives the correct answer for this projection. However, it should be noted $\Delta L$ values of $\sim 3-4$ are still visible, indicating that at a very low level the subhalo is still improving the fit to the data.

\subsection{Subhalo test results}
\begin{figure}
	\includegraphics[width=1.0\columnwidth]{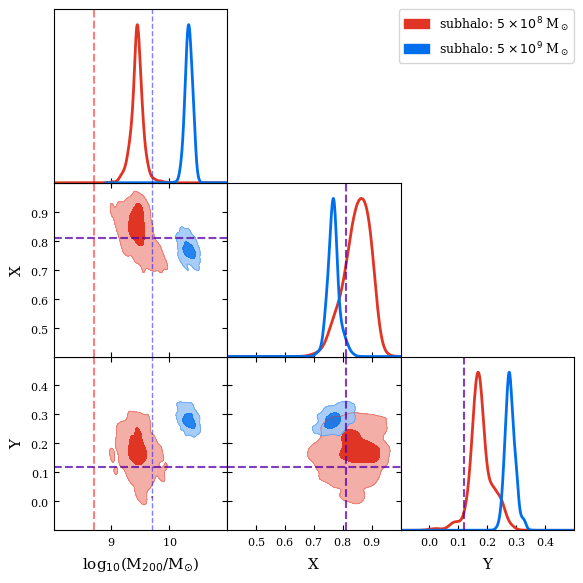}
    \caption{Posterior probability distribution of the subhalo model parameters, after the subhalo refining phase, when modelling the Projection 1 lens mass with an eBPL. Red and blue correspond to the cases of a $5\times10^8$ and $5\times10^9$~M$_\odot$ input subhalo respectively. The 2D contours cover the 68\% and 99\% credible regions. For 1D posteriors, the vertical dashed lines mark the true input values.}
    \label{fig:posteriors_eBPL_P1}
\end{figure}

\begin{figure}
	\includegraphics[width=1.0\columnwidth]{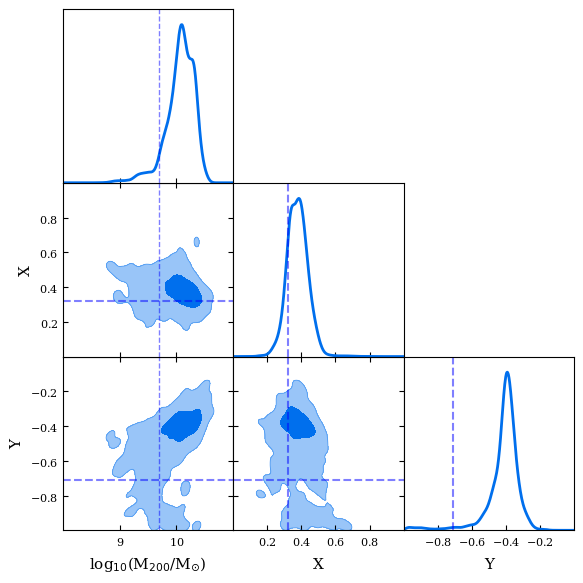}
    \caption{Subhalo parameters' posteriors of the subhalo refining phase, when modelling the Projection 2 lens mass with an eBPL. Only the case of a $5\times10^9$~M$_\odot$ input subhalo is shown (the case of a $5\times10^8$~M$_\odot$ subhalo does not lead to a clear detection). The 2D contours cover the 68\% and 99\% confidence regions. For 1D posteriors, the vertical dashed lines mark the true input values.}
    \label{fig:posteriors_eBPL_P2}
\end{figure}

Having shown the performance of using an eBPL to fit images without a subhalo present, we now test whether the same pipeline can correctly recover a subhalo's properties when a subhalo is included when generating the mock data. For both projections, we add an NFW-like subhalo of $m_{\rm 200} = 5\times10^8$\,M$_\odot$ or $5\times10^9$\,M$_\odot$ at the positions marked by the red crosses in Fig.~\ref{fig:cvgc_mocks}. 

\begin{figure*}
	\includegraphics[width=2.0\columnwidth]{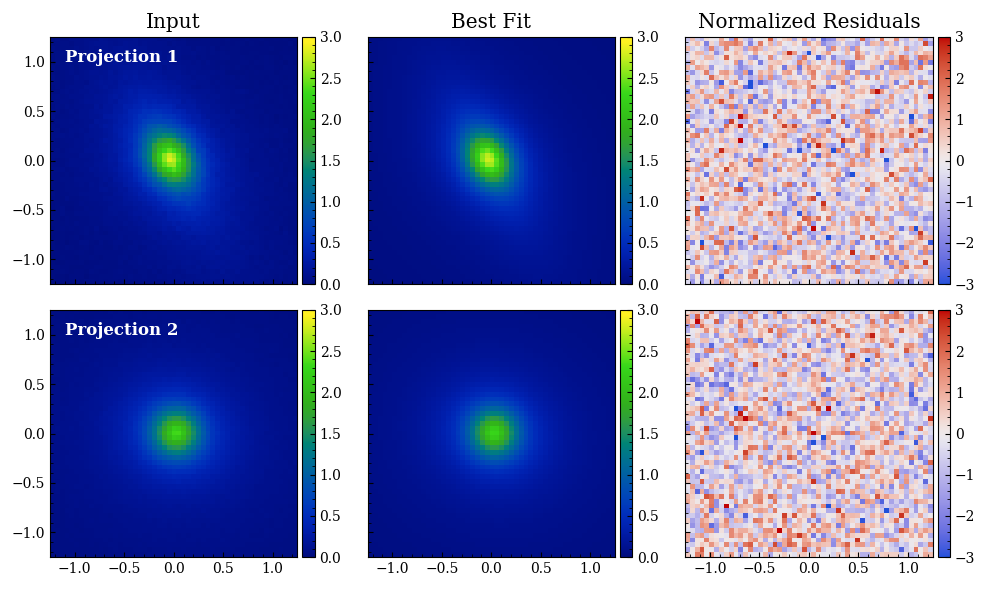}
    \caption{\textbf{Left column:} the mock lens light images. \textbf{Middle column:} the best recovered images using three cored S\'ersics to model the lens light. \textbf{Right column:} the corresponding normalized residuals. The top row show the case of Projection 1 and the bottom row show the case of Projection 2. The units of the y and x axes are arcsec.}
    \label{fig:light_3CS}
\end{figure*}

\begin{figure*}
	\includegraphics[width=2.0\columnwidth]{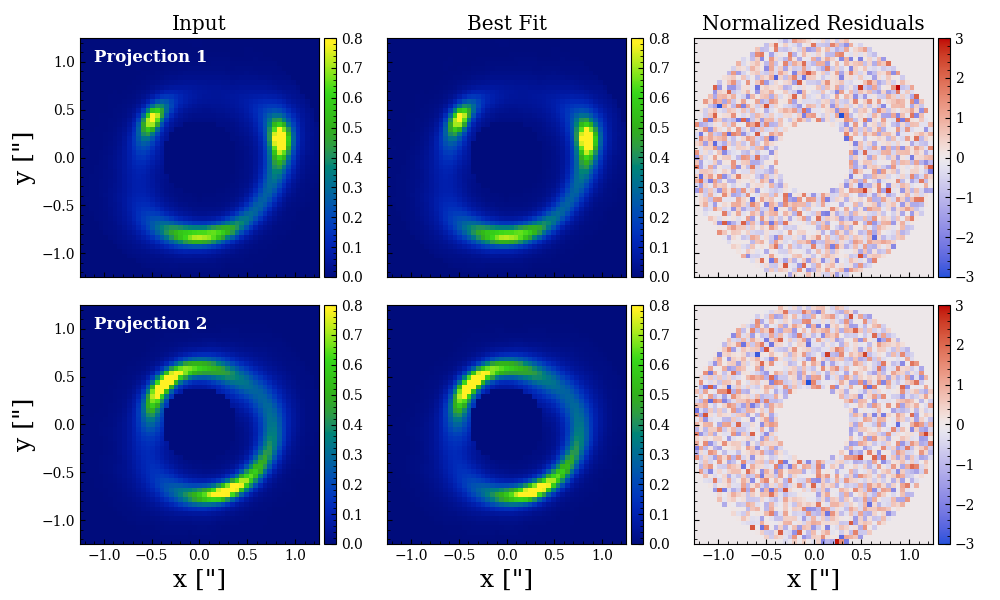}
    \caption{\textbf{Left column:} the mock lensing images. \textbf{Middle column:} the best recovered images using the decomposed model to fit the lens mass. \textbf{Right column:} the corresponding normalized residuals (residuals divided by the noise). The top row show the case of Projection 1 and the bottom row show the case of Projection 2. The colour bars for the left two columns are in units of e$^{-}$~pix$^{-1}$~s$^{-1}$.}
    \label{fig:fit_images_3CS}
\end{figure*}

Similar to our earlier analysis, we first check the $\Delta L$ maps. The middle column of Fig.~\ref{fig:pl_smth_lklhd} shows $\Delta L$ maps for the cases where a $5\times10^8$~M$_\odot$ subhalo is added and the right panels show the results for a $5\times10^9$~M$_\odot$ subhalo. The upper and lower panels show the results of Projection 1 and 2 respectively. For Projection 1, grid cells near the subhalo's true input location (marked as white triangles in the image) show clear increases in  $\Delta L$. For the $5\times10^8$~M$_\odot$ subhalo case, the maximum $\Delta L$ is $\sim$ 24.0 (for the rightmost cell on the fourth row from bottom) and for the $5\times10^9$~M$_\odot$ case, the maximum $\Delta L$ is $\sim$ 81.4 (also for the rightmost cell on the fourth row from bottom). Note that the colourbar saturates (for cells with $\Delta L > 40$) in the top-right panel. 

Based on the detections shown in the $\Delta L$ maps, we continue the subhalo pipeline and fit a model where we no longer confine the subhalo within a particular square cell, and instead use a 2D Gaussian prior on the subhalo position, centered on the best-fit position from the ``grid-search'' phase, with a standard deviation of 0.5\arcsec. In Fig.~\ref{fig:posteriors_eBPL_P1}, we show the posterior of the subhalo parameters for both the case with a $5\times10^8$~M$_\odot$ subhalo (red) and a $5\times10^9$~M$_\odot$ subhalo (blue). The true input values are marked by the dashed lines. As shown, for both cases, the subhalo's mass is significantly overestimated and the true input subhalo masses are excluded by 99\% confidence regions. When the input subhalo has a mass of $5\times10^8$~M$_\odot$, the recovered subhalo mass is overestimated by $\sim$ 5 times with a value of $10^{9.4^{+0.4}_{-0.4}}$~M$_\odot$ inferred, whereas for the input subhalo with $5\times10^9$~M$_\odot$, the recovered mass is overestimated by around 4 times and has a value of $10^{10.3^{+0.2}_{-0.2}}$~M$_\odot$.

For Projection 2, with an input subhalo of $5\times10^8$~M$_\odot$, the maximum $\Delta L$ is only 3.4, therefore no subhalo is detected and we do not analyse the posterior on the subhalo properties. When the input subhalo mass is $5\times10^9$~M$_\odot$, the maximum $\Delta L$ is 9.7  providing us with a plausible detection. We take this plausible detection and refine the fit, with the resulting posterior for the subhalo parameters plotted in Fig.~\ref{fig:posteriors_eBPL_P2} . For Projection 2, the subhalo's mass is recovered to be $10^{10.1^{+0.5}_{-1.2}}$~M$_\odot$ where the errors mark the 99\% confidence regions. It is noted that in Fig.~\ref{fig:posteriors_eBPL_P2}, although the subhalo's true parameters are recovered within 99\% confidence regions (the light blue regions), the best-fit $m_{200}$ and y coordinate are clearly offset from the true input and in a case of smaller errors (higher S/N images), the true inputs might be ruled out.

\subsection{Parametric Source}

To verify that our conclusions are not a result of a systematic associated with our pixelised source model, in appendix \ref{AppendixA} we re-perform all of the above fits assuming a cored S\'ersic profile for the source. For Projection 1 we see nearly identical behaviour in terms of false positives and the subhalo inference, however the $\Delta L$ values are much larger; of order $\sim 250$ compared to the values of $\sim 25$ seen for the pixelised source. This is expected, as the greater flexibility of the pixelised source reduces our sensitivity to a subhalo and therefore also false positives \citep{Gilman2020b}. For Projection 2, fits to the smooth data now infer a false positive with $\Delta L = 18.1$. This does not contradict the results using a pixelised source above, instead the values of $\Delta L = \sim 3-4$ shown in Fig.~\ref{fig:pl_smth_lklhd} have simply been boosted above our threshold value of $\Delta L = 5$ because fitting a cored S\'ersic increases our sensitivity to subhaloes (and false positives). Thus, the eBPL does still produce false positive detections when it fits Projection 2, however to see these using a pixelised source one would likely require much higher S/N data.

\subsection{Summary}

For Projection 1, we saw false positive detections and an inability to recover an input subhalo's mass correctly. In contrast, for Projection 2, we did not infer a false positive detection (when assuming a pixelised source) and the subhalo's true mass is covered by the posterior. Taking into account the different convergence shapes of the two projections (see figure \ref{fig:cvgc_mocks}), we speculate the inaccurate inferences on subhaloes for Projection 1 are caused by the clear mismatch in the shape of the eBPL and the more elliptical input profile. We also speculate the better performance seen for Projection 2 is because its rounder convergence is easier for the eBPL to model. However, due to the limited number of projections available, we cannot generalize these conclusions any further. We only saw the eBPL produced false positives in Projection 1 and there is a possibility that it is a different property of the lens driving this result. When analysing real lens systems we will look to see whether departures from ellipticity in the lens galaxy's light (Nightingale et al. 2022, in prep.) are correlated with subhalo detections, possibility indicating a false positive signal.

\section{A DECOMPOSED Model}\label{sec:star_dark}

Motivated by the inability of the eBPL to provide a robust subhalo inference, we now consider the decomposed model, which models a galaxy's stellar and dark matter mass separately. This includes sufficient freedom to capture complex features such as a pointy convergence profile, or other departures from elliptical symmetry.

\subsection{Model introduction}
In most strong lens images, we observe not only the lensed source's light, but also the light emitted from the lens galaxy, which should approximately trace its stellar mass distribution. For example, through inspection of the lens galaxy's light profile, we can estimate the position angle and axis ratio of the lens's stellar mass profile. More detailed light profile fits can provide us with a more detailed model of the stellar mass distribution. We now explore the potential of utilizing this information and if it can allow us to correctly recover the subhalo information hidden in the source's lensed images. We fit the lens's mass using a decomposed model which treats the lens galaxy's stellar mass and its dark matter mass separately. This type of model has been fitted in many previous studies \citep{Dye2005, Suyu2014, Wong2017} and \citet{Nightingale2019} showed using HST imaging of three SLACS lenses that such models capture variations in ellipticity and position angle within a galaxy that are indicative of pointy mass distributions.

For the stellar mass, we assume it exactly traces the stellar light, which allows us to directly transform between the two by multiplying by a constant mass-to-light ratio (M/L) parameter, which can be described as
\begin{equation}
\label{eqn:Sersickap}
\kappa (r) = \Psi\cdot I(r) \, \, ,
\end{equation}
where $I(r)$ corresponds to the light profile and $\Psi$ is its ``mass-to-light ratio''. For simplicity, we directly take the input stellar mass of the simulation's particle data as our lens light and therefore do not consider a more realistic galaxy light simulation process. In that sense, the ``$I(r)$'' is equivalent to the convergence profile and thus $\Psi$ becomes a dimensionless quantity and is set to be 1.0. 

To utilize the ``lens light'' information, we model the ``lens light`` with three cored S\'ersic profiles as described by Eq.~\ref{equ:core_sersic}. We opt for the cored S\'ersic because of the simulated galaxy's core; for real lenses we anticipate that the regular non-cored S\'ersic profile will suffice. We impose that the 3 cored S\'ersic profiles share the same centre, but allow for them to have different position angles and axis ratios. We use three profiles because fits using two profiles do not fully capture the features of the ``lens light'' (e.g. clear spatially-correlated normalised residuals are seen when the best-fit two cored S\'ersic model is subtracted from the true stellar mass distribution). In Fig.~\ref{fig:light_3CS}, we show the input lens light (left column), best-fit 3 cored S\'ersic profiles (middle column) and corresponding normalised residuals (right column). For both projections (upper row corresponds to the Projection 1 and the lower one is for the Projection 2) the light is well fit by three cored S\'ersics. Later, in our lens mass modelling, we fix the stellar mass distribution to be exactly the same as the best-fit three cored S\'ersic profiles obtained from fitting the lens light, except for a free $\Psi$ which changes the overall normalisation of the projected stellar mass distribution.

In addition to the stellar mass, we include an elliptical NFW profile into the lens model (to account for the dark matter). This has six free parameters: a scale radius, $r_{\rm s}$, and scale convergence, $\kappa_{\rm NFW}$; two ellipticity components; and the 2D coordinates of the halo centre. As in the eBPL case, we include an external shear in the decomposed lens model.

Neither cored S\'ersic nor elliptical NFW profiles have analytical formulae for their deflection angles. For fast computation we follow \citet{Shajib2019} and use a sum of 2D Gaussian profiles to approximate the cored S\'ersic and elliptical NFW profiles. The resulting deflection angles are simply a sum of the deflection angles of the individual Gaussian profiles, which can be efficiently computed using analytical formulae. To be specific, in our work, in most cases we approximate a cored S\'ersic profile by 30 Gaussian profiles with their standard deviations uniformly distributed in the log$_{10}$ space between $0.01r_{\rm e}$ and $50r_{\rm e}$, where $r_{\rm e}$ is the effective radius of the cored S\'ersic. Similarly, for an elliptical NFW profile, we also approximate it with 30 Gaussians and the standard deviations of those Gaussians are uniformly distributed in the log$_{10}$ space between $0.0005r_{\rm s}$ and $30r_{\rm s}$. We noticed that one of the best-fit cored S\'ersic components to the ``lens light'' of Projection 2 has a S\'ersic index of 0.51 and for that profile the decomposition formula (Eq.~5 of \citet{Shajib2019}) becomes numerically unstable. For that one particular case, we instead decompose the S\'ersic profile into a sum of Gaussians using \citet{Cappellari2002}'s method, which optimizes the standard deviations and amplitudes of those Gaussians at the same time. We have tested our choices of the parameters of the Gaussian decomposition method across a large variety of cored S\'ersics and elliptical NFW profles to ensure that errors of approximating the deflection angles are much smaller than the perturbation of a subhalo of interest. In Table~\ref{tab:3CS}, we summarize our lens model parameters.  

The approach we follow cannot be straightforwardly translated to real data. For example, we have modelled the lens's light in the absence of the source light and ignored potential complications such as a radial gradient in the mass-to-light ratio. The goal of this work is not to present a method that can be directly transferred to the fitting of real data, but simply to show that when sufficient complexity is added to the lens mass model one's inference on subhalo properties improves. Nevertheless, \citet{Nightingale2019} have already shown how {\tt PyAutoLens} can fit this type of model to real data and we expand on this further in section \ref{sec:applied_to_data}.

\begin{table}
    \renewcommand\arraystretch{1.5}
	\centering
	\begin{tabular}{ccc} 
	\hline
	& \textbf{Projection 1} & \textbf{Projection 2} \\
	\hline
	\textbf{Stellar Mass} & \multicolumn{2}{c}{\textbf{3 Core Sersics}} \\
	\hline
	 centre(x, y) [(\arcsec, \arcsec)] & (0.008, -0.036) & (0.003, 0.022) \\
     $I^{'}$ & \{0.44, 0.60, 0.31\} & \{0.32, 0.50, 1.06\} \\
	 $r_{\rm e}$ [\arcsec] & \{0.65, 0.02, 4.42\} & \{0.11, 0.18, 2.39\} \\
	 $r_{\rm c}$ [\arcsec] & \{0.14, 0.27, 0.25\} & \{0.35, 0.13, 0.02\} \\
	 $n$ & \{1.44, 4.36, 4.91\} & \{2.64, 0.51, 2.31\} \\
	 position angle [$^{\circ}$] & \{-62, -59, -45\} & \{-27, 69, -73\} \\
	 axis ratio & \{0.33, 0.82, 0.82\} & \{0.90, 0.89, 0.90\} \\
	 $\Psi$ & \multicolumn{2}{c}{[0.8, 1.2]} \\
	 redshift & \multicolumn{2}{c}{0.2} \\
	 MGE \{n, rmin, rmax\} & \multicolumn{2}{c}{\{30, $0.01r_{\rm e}$, $50r_{\rm e}$\}} \\
	\hline
	\textbf{Dark Matter Mass} & \multicolumn{2}{c}{\textbf{NFW}} \\
	\hline
	centre($x$, $y$) [(\arcsec, \arcsec)] & \multicolumn{2}{c}{[-0.1, 0.1]} \\
	log$_{10}$ $\kappa_{\rm NFW}$ & \multicolumn{2}{c}{[-2, 0.3]} \\
	$r_{\rm s}$ [\arcsec] & \multicolumn{2}{c}{[10, 50]} \\
	$e_{1}$ & \multicolumn{2}{c}{[-1.0, 1.0]} \\
	$e_{2}$ & \multicolumn{2}{c}{[-1.0, 1.0]} \\
	redshift & \multicolumn{2}{c}{0.2} \\
	MGE \{n, rmin, rmax\} & \multicolumn{2}{c}{\{30, $0.0005r_{\rm s}$, $30r_{\rm s}$\}} \\
	\hline
	\textbf{External Shear} & & \\
	\hline
	$\gamma_{\rm 1ext}$ & \multicolumn{2}{c}{[-0.2, 0.2]} \\
	$\gamma_{\rm 2ext}$ & \multicolumn{2}{c}{[-0.2, 0.2]} \\
	\hline
	\textbf{Subhalo} & \multicolumn{2}{c}{\textbf{Spherical NFW}} \\
	centre($x$, $y$) [(\arcsec, \arcsec)] & \multicolumn{2}{c}{([-1.0, 1.0], [-1.0, 1.0])} \\
	log$_{10}$ $m_{\rm 200}$ [M$_\odot$] & \multicolumn{2}{c}{[6, 11]} \\
	mass-concentration relation & \multicolumn{2}{c}{\citet{Ludlow2016}}
	\\
	\hline
	\end{tabular}
	\caption{Parameters and priors for the decomposed model. Parameters with values shown in ``()'' or ``\{\}'' are fixed during the modelling. Parameters with values shown as ``[a, b]'' are fit for, with a uniform prior between a and b.}\label{tab:3CS} 
\end{table}

\subsection{Results}

We now present results using the decomposed model, following the same structure we used for the eBPL results, whereby we begin with the smooth test results (where no subhalo is present in the simulated data) followed by results where the simulated data includes a subhalo.

\begin{figure*}
	\includegraphics[width=2.0\columnwidth]{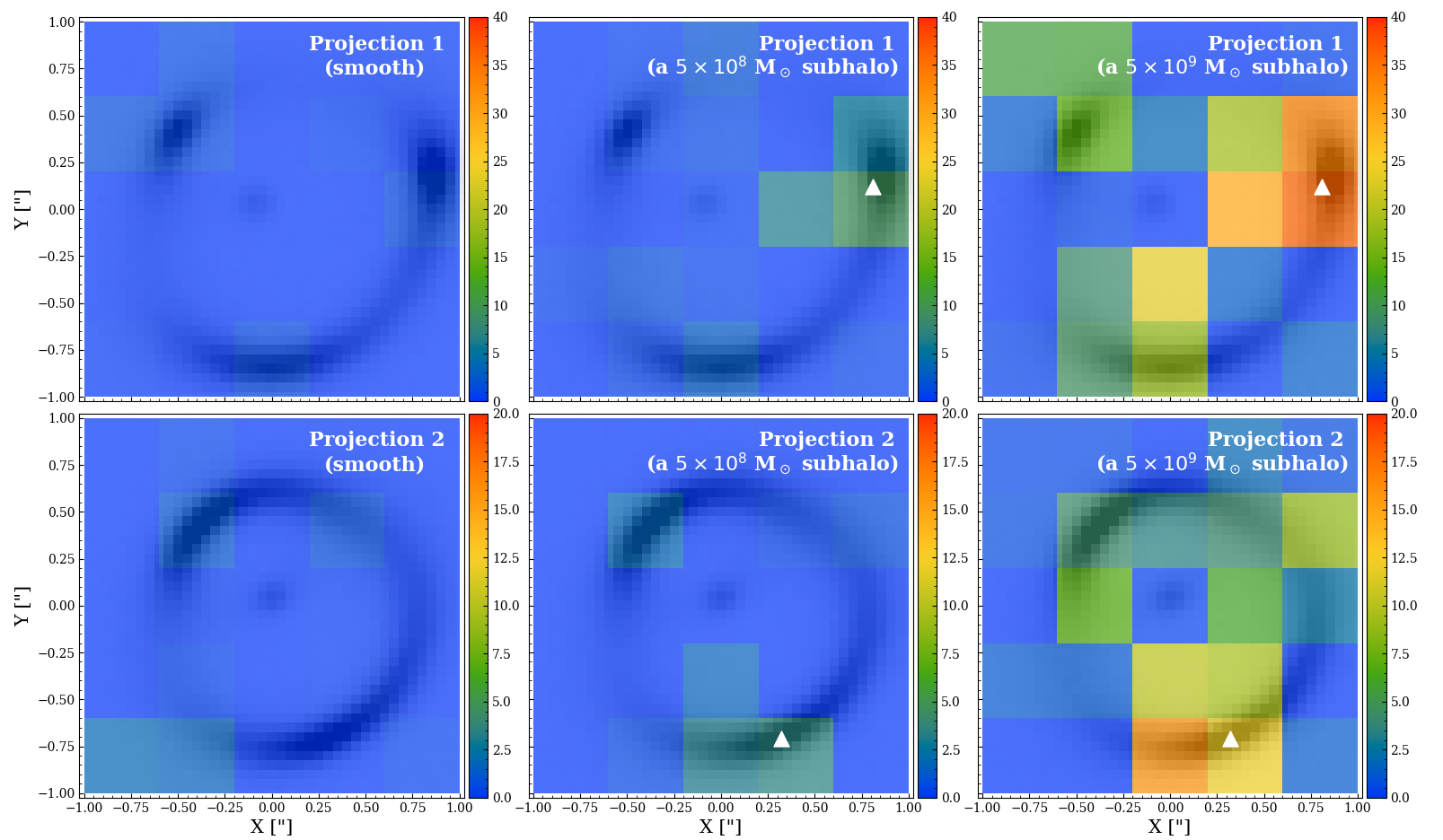}
    \caption{Modelling the lens with a decomposed stellar $+$ dark matter model removes false-positive detections, and yields correct subhalo masses. Colours indicate the increase in maximum log likelihood, $\Delta L$, when a subhalo is included in the fit to a lens that: has no subhaloes (left), has a subhalo of mass $5\times10^8$~M$_\odot$ (middle), or has a subhalo of mass $5\times10^9$~M$_\odot$ (right).
    The position of the subhalo in the fit is free to vary within squares of side $0.4\arcsec$. The top and bottom rows show the results for Projection 1 and 2 (with different colour scales). White triangles mark the true locations of the subhaloes.}
    \label{fig:pl_smth_lklhd_3CS}
\end{figure*}

In Fig.~\ref{fig:fit_images_3CS}, we compare the input and best-fit model images for smooth cases. As shown by the normalized residuals in the third column, no clear correlated residuals exist, which indicates an overall good fit with the decomposed model. Comparing the results with the equivalent BPL results in Fig.~\ref{fig:fit_images}, we see that the BPL results are indistinguishable from the decomposed model results in terms of the residuals, which confirms again that ``subhalo-like'' perturbations cannot be detected visually from the residual maps and we have to rely on careful statistical comparisons to make inferences about subhaloes.

In the left column of Fig.~\ref{fig:pl_smth_lklhd_3CS}, we first show the maximum log likelihood difference maps when modelling the smooth image with the decomposed model described above. For both projections the decomposed model fits the image accurately with a maximum $\Delta L$ value below 5, correctly indicating that no subhalo exists in the lens galaxy. Unlike the eBPL, the decomposed model does not give false-positive signals in our ``smooth tests''. 

\begin{figure}
	\includegraphics[width=1.0\columnwidth]{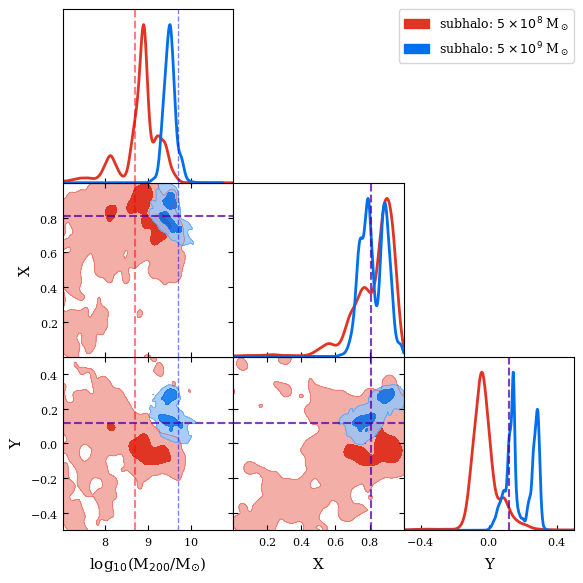}
    \caption{The posteriors on the subhalo parameters from the subhalo refining phase, fitting the decomposed model to mock data generated using Projection 1. Red and blue colours show the cases with an input subhalo mass of $5\times10^8$ and $5\times10^9$~M$_\odot$ respectively. The 2D contours cover the 68\% and 99\% confidence regions. For 1D posteriors, the vertical dashed lines mark the true input values.}
    \label{fig:posteriors_3CS_P1}
\end{figure}

In the middle and right columns of Fig.~\ref{fig:pl_smth_lklhd_3CS}, we show the $\Delta L$ maps when a $5\times10^8$~M$_\odot$ or $5\times10^9$~M$_\odot$ subhalo is added to the lens galaxy at the positions marked by the white triangles. For Projection 1 (upper panels), the regions where we detect the maximum $\Delta L$ is consistent with the position of each input subhalo. For an input subhalo of $5\times10^8$~M$_\odot$, the result shows a plausible detection where the maximum $\Delta L$ is $9.3$, whereas for an input subhalo of mass $5\times10^9$~M$_\odot$, the detection is even clearer with a maximum $\Delta L$ of $36.0$. Having successfully detected the subhalo in each case, we continue on to the subhalo refining fit, with Fig.~\ref{fig:posteriors_3CS_P1} showing the inferred posteriors of the subhalo parameters. For both cases, the subhalo parameters are correctly recovered within 99\% credible regions. For a $5\times10^8$~M$_\odot$ subhalo, the recovered value is $10^{8.9^{+0.8}_{-2.6}}$~M$_\odot$, and for a $5\times10^9$~M$\odot$ subhalo, the recovered value is $10^{9.5^{+0.5}_{-0.4}}$~M$_\odot$.

\begin{figure}
	\includegraphics[width=1.0\columnwidth]{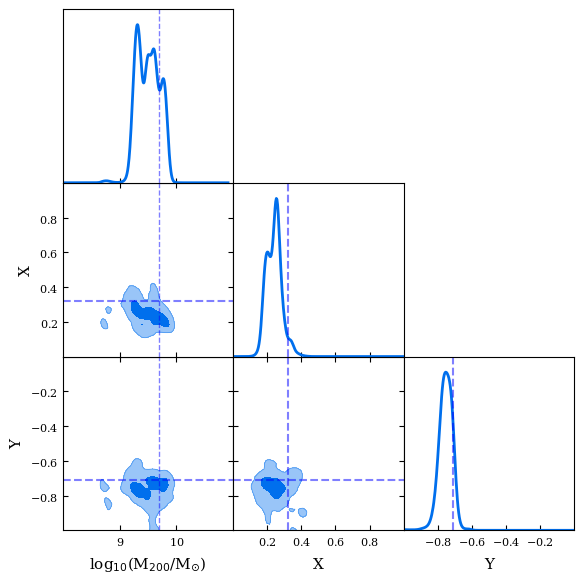}
    \caption{The same as Fig.~\ref{fig:posteriors_3CS_P1}, but for Projection 2. Only the case with a $5\times10^9$~M$_\odot$ subhalo is shown (a $5\times10^8$~M$_\odot$ subhalo is not clearly detected).}
    \label{fig:posteriors_3CS_P2}
\end{figure}

For Projection 2, we only get a detection when the true subhalo mass is $5\times10^9$~M$_\odot$, with a maximum $\Delta L$ of 15.8 (in the middle cell of the bottom row). With a $5\times10^8$~M$_\odot$ subhalo, all subhalo-position cells have $\Delta L < 5$, corresponding to no detection. In Fig.~\ref{fig:posteriors_3CS_P2}, we show the subhalo posteriors obtained from the subhalo refining phase for the $5\times10^9$~M$_\odot$ case. We recover the input subhalo mass, with a 99\% credible region on $m_{\rm 200}$ of $10^{9.5^{+0.4}_{-0.7}}$~M$_\odot$. For the non-detection of the $5\times10^8$~M$_\odot$ subhalo, we do not believe this is a failure of the decomposed model, but instead a limitation of the data quality. In fact, if we check the inferred subhalo parameters for the sub grid cell which contains the input subhalo, the inferred subhalo's mass is $10^{8.7^{+0.7}_{-2.5}}$~M$_\odot$, which is still consistent with the true input mass. Thus, our inference on the subhalo's parameters is consistent with the truth, but we have insufficient S/N for the model to be favoured in terms of $\Delta L$.

We note that in Fig.~\ref{fig:posteriors_3CS_P1} and \ref{fig:posteriors_3CS_P2}, the posterior distributions are not smooth, having a ``patchy'' appearance in the 2D marginalised posteriors and ``wiggles'' in the 1D posteriors. These arise due to the pixelised source plane. The source plane pixelisation is created from a Voronoi tessellation of generating points, where the generating points are first placed in the image plane and then mapped into the source plane. Changes to the mass model change the mapping from image plane back to the source plane, such that the positions and shapes of the source-plane pixels varies as the lens mass model is changed. Certain locations for pixel boundaries may be more or less able to reproduce the observed data, leading to small changes to the mass model parameters capable of macroscopic changes to the likelihood. This phenomenon is more significant for more complex mass models which have more parameters and freedom to allocate those source pixel grids on the source plane. As a result, we get unsmooth posteriors for our decomposed model. The work of \citet{Etherington2022} discusses this further and presents a solution using a cap on the log likelihood of the model-fit.

\subsection{Parametric Source}

In appendix \ref{AppendixA}, we again verify that our conclusions hold when we assume a cored S\'ersic profile for the source. For both projections, the decomposed model does not give a false positive; with the highest value of $\Delta L = 2.5$. Note that, for a cored S\'ersic source, false positives were detected for both projections. Given that fits assuming a cored S\'ersic for the source give a much higher sensitivity to subhaloes and false positives, this further strengthens our conclusion that by adding the right type of complexity to the decomposed mass model removes the presence of false positives. As a result of this increased sensitivity, the $5\times10^8$~M$_\odot$ subhalo is also detected successfully in Projection 2, which is not the case for the pixelised source.

\subsection{Offset True Positive Detections}

In the right panels of Fig.~\ref{fig:pl_smth_lklhd_3CS} we note increases of $\Delta L = \sim 10-15$ away from the true location of the $m_{200}=5\times10^{9}$~M$_\odot$ subhalo for both projections. These are solutions where an offset dark matter subhalo closely mimics the perturbing effects of the actual subhalo in the data. However, it is not a perfect representation of the actual subhalo, which is why fits at the true location infer higher overall log likelihood values. We do not consider these as false-positive detections because they are caused by the true presence of a detectable subhalo in the data. Should this behaviour be seen in real data we therefore should not discount the signal as a false positive. In fact, a candidate subhalo detection should be made and followed up with a second subhalo search which includes the first subhalo in the model, so as to validate the detection.

\subsection{Summary}

For both projections the decomposed model is a success. When we do not include a subhalo in the input lens galaxy, it returns no detections. When a subhalo is included in the mock data, it is able to correctly infer the existence of the subhalo through an increase in $\Delta L$ (at least for three out of the four cases we tried). Furthermore, it recovers the masses and positions of the subhaloes within 99\% credible regions. By utilizing (idealized) ``lens light'' information, the decomposed model therefore successfully captures complexity in the mass profile (e.g. the non-elliptical shape) that the eBPL could not. The success of the decomposed model confirms that for subhalo detection, it is vital to model the lens galaxy's mass accurately.

\section{Discussion}\label{sec:discussion}

\subsection{Implications for strong lensing subhalo detection}

Our results confirm that if a dark matter subhalo is located near the emission of a strongly lensed source galaxy, its perturbing effects mean that its presence can be inferred. For gravitational imaging\footnote{Similar tests on flux ratios have previously been explored by \citet{Hsueh2018}.}, our work demonstrates this  for first time by simulating the lens galaxy using a mass distribution derived from the particle data of a cosmological simulation, which therefore does not make idealized assumptions like a single axis of ellipticity. However, we also showed that assuming an overly simplistic mass model for the lens galaxy which lacks certain complexity compared to the true underlying mass distribution has two negative effects on the subhalo inference: (i) it may lead to false-positive detections of a dark matter subhalo even though a subhalo is not present in the data and; (ii) when a subhalo is truly present in the data it may lead to systematic biases on the inferred subhalo mass by a factor of $4-5$.

The notion that a mismatch in mass profile shape could lead to false-positive subhalo detections supports the analysis of \citet{Ritondale2019}, who noted several false-positive signals found in real lensing systems in the BELLS-GALLERY sample. For example, they noted an increase in log Bayesian evidence of 72 in the lens SDSSJ0755+3445, but demonstrated -- using a potential correction technique \citep{Koopmans2005, Vegetti2009b} -- that the mass model could be improved by small corrections over a large angular scale, as opposed to a localised correction reminiscent of a subhalo. This indicates that the subhalo-like signal is probably due to the mismatch in the macro models, as we saw in our tests. False positives are also partly the reason why \citet{Vegetti2014} and \citet{Vegetti2018} require Bayesian evidence increases of $50$ and $100$ to claim a dark matter detection; values below this threshold may be false positives (the authors also require validation via potential corrections). Whilst the false positives in this work did not create Bayesian evidence increases above $30$, the overall size of the increase depends on the properties of the strong lens and sources simulated, the S/N of the data and model used to fit the data. In appendix \ref{AppendixA} false positives with evidence increases above $200$ are inferred. Therefore, \textbf{our results do not indicate that previous detections of dark matter subhaloes in strong lenses are false positives}. Instead, they show the importance of techniques like the potential corrections and we provide insight on why these methods are able to distinguish a subhalo detection from missing complexity in the mass model. 

\subsection{Is our simulated lens galaxy realistic?}

It is important to consider how realistic the simulated galaxy used in this work is. As discussed previously, the galaxy was selected to be similar to lens galaxies from the SLACS survey. It has a typical halo mass for a SLACS lens and its stellar mass and size (i.e. the half-light radius) follow recent observations of massive galaxies \citep{Huang2018}. The central galaxy has a complex shape, with isophotes that change shape when viewed from different directions, and where the shapes of the isophotes can vary with radius for a fixed viewing direction. We speculated that this departure from elliptical symmetry drove the false positives, because they are only seen for the projection where the mass distribution is highly elliptical. A varying isophotal shape with radius is commonly seen in observations of massive elliptical galaxies with comparable mass to SLACS strong lenses. For example, over 1/3 of galaxies with stellar masses above $10^{11.5}$~M$_\odot$ taken from the MASSIVE survey show isophotal position-angle rotations \citep{Goullaud2018}, known as ``isophotal twists'' (see \citet{Oh2017} for similar results in lower mass early-type galaxies). Similar features are also reported in three strong lenses by \citet{Nightingale2019}. We therefore believe this aspect of our simulation is representative of real strong lenses and is a plausible cause of some of the false positives in the SLACS and BELLS-GALLERY lenses discussed previously.

The simulated galaxy also has a sub-kpc core, which generates a central image in our mock lens images. This phenomenon is seen in other works which simulate strong lenses from cosmological simulations \citep{Mukherjee2018, Despali2020, Ding2021}, with the core due to insufficient simulation resolution. Central images of this brightness are not seen in real observations of strong lenses, therefore such a large core is unrealistic. To ensure it does not impact our tests, the mass model parameterizations fitted in this work all included cores. We masked the central image so as to ensure the mass models did not utilize additional information that is not present in real images of strong lenses. Whilst this aspect of the simulated lens is therefore not realistic, the mass modeling performed in this work ensures we can generalize our conclusions to the analysis of real data.

\subsection{Application to real data}\label{sec:applied_to_data}

Our next step is applying the decomposed model to real data. We expect that we will be able to fit mass models which omit parameters that account for a core, given that the core feature is a consequence of the inadequate simulation resolution. For the decomposed model, we will likely fit regular S\'ersic functions instead of the cored S\'ersics fitted in this work.

The decomposed model verified that if a mass model can accurately capture the lens galaxy's complexity, it will improve the subhalo inference. This work used information from the simulation that is not available when analysing real data, for example we utilized our true knowledge of the lens's stellar mass distribution. Nevertheless, we believe these models can be translated to real data, where the light emitted from the lens galaxy acts as a tracer for the stellar mass, information which is often omitted when modeling a strong lens (e.g. by assuming a power-law mass model). This approach to lens modeling was explored in \citet{Nightingale2019}, who fitted a decomposed stellar plus dark matter to three strong lenses. The authors showed that all three lenses showed isophotal twists in their stellar emission and that when this was modelled using two stellar components with different ellipticities and position angles it improved the mass model compared to a model assuming a single elliptical geometry. We are now investigating whether these lens systems produce subhalo detections, which would be indicative of a false positive.

The decomposed model must also make assumptions in converting light to mass. For example, whether the S\'ersic profiles representing each stellar component share the same mass-to-light ratio or whether each ratio is a free parameter in the model. For each component, one must also choose whether the lens model accounts for a radially varying M/L \citep{Napolitano2005, Tortora2011, Ge2021}. The assumption of an elliptical NFW profile to describe the dark matter poses another possible mismatch. The main concern on small scales is whether the central slope, which in simulations is affected by the presence of baryons, is equal to the NFW one. To take this into account when modelling real data, we could model the dark matter as a profile with a free central slope, e.g. a generalized NFW profile \citep{Zhao1996}, or explicitly model the way baryons are expected to alter the dark matter distribution \citep{Cautun2020, Callingham2020}. We do not expect this to be a significant issue since for galaxy-galaxy strong lensing, the dark matter mass is typically sub-dominant in the region of interest \citep[e.g.][]{Hongyu2016, Li2019}. In future work we will seek to understand the importance of all these different assumptions with a view to improving the dark matter subhalo inference.

\subsection{Subhalo Sensitivity}

If the decomposed model can be successfully fitted to real data, it also has implications for how sensitive strong lensing is to low mass dark matter subhaloes. Firstly, if the method is able to reduce or remove the Bayesian evidence thresholds applied by works like \citet{Vegetti2014} to remove false positives, this will make us sensitive to lower mass subhaloes (which produce smaller evidence increases). Furthermore, because the decomposed model uses the stellar light as additional information which constrains the mass model, this may further boost one's sensitivity to subhaloes by reducing the degeneracy between the lens galaxy's mass model and subhalo. This will require that sensitivity mapping of a strong lens, which quantifies what mass subhaloes one will detect if truly present in the data \citep{Despali2020, He2021, Amorisco2022}, is performed using the decomposed model, as opposed to the power law model assumed in previous studies. The same level of care will be necessary in understanding how robust assumptions associated with the M/L and dark matter are.

\subsection{Other Lensing Studies}

A mass model mismatch has also been discussed in the analysis of strongly lensed quasars. \citet{Hsueh2017} showed that the flux ratio anomalies observed in lens system CLASS B0712+472 can be largely resolved by additionally adding a disk profile to the lensing model. The works of \citet{Gomer2020, Gomer2021, Cao2021, Van2021} show that such mismatches can impact on the inference of the Hubble constant via time-delay cosmology.

\section{Conclusions}\label{sec:conclusions}

With a large increase in the number of observed galaxy-galaxy strong lenses expected within this decade, strong lensing could soon push the constraints on the halo mass function to low enough masses that it provides evidence in favour of or against warm dark matter models. However, detecting dark matter subhaloes through strong lensing is a challenging problem due to the complexity of the lens galaxy's mass distribution. In this work, we use a massive elliptical galaxy extracted from a state-of-the-art hydrodynamic simulation to create mock strong lens images. We represent the simulated galaxy's projected mass distribution as a sum of elliptical Gaussian profiles, which shows departures from the idealized elliptically symmetric mass models typically employed to analyse strong lenses (e.g. the power law profile \citep{Tessore2015}). We project the same simulated galaxy along two different line-of-sight directions, with one projection producing a pointy ``American football like''-shape and the other one appearing rounder.

For each projection, we simulate three strong lens imaging datasets. The first dataset does not include a dark matter subhalo, whereas the other two include a $m_{200}=5\times10^{8}$~M$_\odot$ and $m_{200}=5\times10^{9}$~M$_\odot$ dark matter subhalo near the lensed source's light. To every dataset, we fit two lens mass models: (i) an elliptical broken power law (eBPL) mass model \citep{Oriordan2019} which represents the overall mass distribution of the lens galaxy (e.g. stars and dark matter) and; (ii) a decomposed model that models the stellar and dark matter mass separately (using the stellar particle data from the simulation to constrain part of the stellar mass model). For both models, we investigate fits which include a dark matter subhalo in the lens mass model, and therefore quantify whether we can accurately recover a dark matter subhalo when it is included in the simulation as well as whether we incorrectly infer the presence of a subhalo when it is not truly there; a false positive. 

Our main results can be summarized as follows:

\begin{itemize}
    \item When using an eBPL model to fit the lens mass to the pointy projection without a dark matter subhalo, a false-positive detection is inferred at over 3$\sigma$ confidence. For the same projection, when a $5\times10^8$~M$_\odot$ or $5\times10^9$~M$_\odot$ subhalo is added to the mock lens, the fit correctly recovers the subhalo but overestimates its mass by a factor of $4-5$, with the true input mass outside the inferred 99\% credible regions. However, when modelling data from the projection with a rounder convergence, the eBPL model does not give a false-positive and recovers the input $5\times10^9$~M$_\odot$ subhalo's mass accurately (the $5\times10^8$~M$_\odot$ subhalo is not detected due to insufficient data quality). 
    
    \item When using the decomposed model to fit the lens mass, for both projections, we get no false positives and correctly recover the properties of an input subhalo when there is sufficient data quality to detect it. 
    
\end{itemize}

The eBPL total mass model therefore shows undesirable results, including false positives and an inaccurate estimate of the subhalo mass, which the decomposed mass model does not. We speculate that this is because the eBPL parameterization does not capture aspects of the simulated lens's mass distribution. In particular, the eBPL does not capture the varying ellipticity and orientation seen in the pointy projection's mass distribution. The decomposed mass model does not assume a single elliptical mass distribution and can therefore account for this variation in ellipticity and orientation. Its improved model of the lens galaxy's mass therefore offers an improved subhalo inference which does not suffer false-positive detections. 

\textbf{Our results do not imply that previous detections of dark matter subhaloes in strong lenses are false positives} (e.g. \citet{Vegetti2014}). These studies are fully aware of the false positive phenomena and they require a subhalo detection to pass stringent criteria to be considered a genuine dark matter subhalo. This includes a pixel-based correction to the gravitational potential \citep{Koopmans2005} which accounts for the types of deficiencies in the mass model discussed in this work. In fact, our work demonstrates that dark matter substructures can be successfully detected in images of strong lenses, even when the lens galaxy's mass distribution is more complex than the mass model assumed to fit it. 

Our work highlights the benefits of using cosmological simulations to test strong lens modeling methodology. When the eBPL showed inaccurate results, we were able to compare directly to the simulation's particle data in order to understand what complexity the model is missing. This is not possible when analysing real images of strong lenses. We are now looking to apply what we have learned in this study to real data, and fit strong lenses from existing lens samples with decomposed mass models which, crucially, relax the assumption of a single axis of ellipticity. Applying the models to real data has challenges, for example instead of relying on the simulation's stellar particle data we will need to use the lens's light to constrain the stellar mass \citep{Nightingale2019}. However, the pay-off could be huge, allowing us to more reliably detect lower mass dark matter substructures, that could potentially push our sensitivity down to pivotal masses of $m_{200}=5\times10^{8}$~M$_\odot$ where many viable alternatives to the cold dark matter model begin to make different, testable predictions.

\section*{Software Citations}

This work uses the following software packages:

\begin{itemize}

\item
\href{https://github.com/astropy/astropy}{{Astropy}}
\citep{astropy1, astropy2}

\item
\href{https://bitbucket.org/bdiemer/colossus/src/master/}{{Colossus}}
\citep{colossus}

\item
\href{https://github.com/dfm/corner.py}{{Corner.py}}
\citep{corner}

\item
\href{https://github.com/joshspeagle/dynesty}{{Dynesty}}
\citep{dynesty}

\item
\href{https://github.com/glenco/glamer}{{GLAMER}}
\citep{Metcalf2014, Pekova2014, Metcalf2020}

\item
\href{https://github.com/matplotlib/matplotlib}{{Matplotlib}}
\citep{matplotlib}

\item
\href{numba` https://github.com/numba/numba}{{Numba}}
\citep{numba}

\item
\href{https://github.com/numpy/numpy}{{NumPy}}
\citep{numpy}

\item
\href{https://github.com/rhayes777/PyAutoFit}{{PyAutoFit}}
\citep{pyautofit}

\item
\href{https://github.com/Jammy2211/PyAutoLens}{{PyAutoLens}}
\citep{Nightingale2015, Nightingale2018, pyautolens}

\item
\href{https://github.com/AshKelly/pyquad}{{Pyquad}}
\citep{pyquad}

\item
\href{https://www.python.org/}{{Python}}
\citep{python}

\item
\href{https://github.com/scikit-image/scikit-image}{{Scikit-image}}
\citep{scikit-image}

\item
\href{https://github.com/scikit-learn/scikit-learn}{{Scikit-learn}}
\citep{scikit-learn}

\item
\href{https://github.com/scipy/scipy}{{Scipy}}
\citep{scipy}

\item
\href{https://www.sqlite.org/index.html}{{SQLite}}
\citep{sqlite}

\end{itemize}

\section*{Acknowledgements}

QH, AA, CSF and SMC acknowledge support from the European Research Council (ERC)  Advanced Investigator grant DMIDAS (GA 786910). JN and RM are supported by STFC via grant ST/T002565/1, and the UK Space Agency via grant ST/W002612/1.
RL and XYC acknowledge support from the National Nature Science Foundation of China (Nos.\ 11988101, 11773032, 12022306), science research grants from the China Manned Space Project (Nos.\ CMS-CSST-2021-B01, CMS-CSST-2021-A01) and support from the K.C.Wong Education Foundation.
AE is supported by STFC via grants ST/R504725/1 and ST/T506047/1. 
This work was performed using the Cambridge Service for Data Driven Discovery (CSD3), part of which is operated by the University of Cambridge Research Computing on behalf of the STFC DiRAC HPC Facility (www.dirac.ac.uk). The DiRAC component of CSD3 was funded by BEIS capital funding via STFC capital grants ST/P002307/1 and ST/R002452/1 and STFC operations grant ST/R00689X/1. DiRAC is part of the National e-Infrastructure. This work also used the DiRAC@Durham facility managed by the Institute for Computational Cosmology on behalf of the STFC DiRAC HPC Facility (www.dirac.ac.uk). The equipment was funded by BEIS capital funding via STFC capital grants ST/P002293/1 and ST/R002371/1, Durham University and STFC operations grant ST/R000832/1. DiRAC is part of the National e-Infrastructure. Additional support was provided by  STFC  grant ST/T000244/1.

\section*{Data Availability}

The data underlying this article will be shared on reasonable request to the corresponding author.



\bibliographystyle{mnras}
\bibliography{example} 



\appendix

\section{Parametric source results}\label{AppendixA}
This work primarily focuses on the lens mass distribution. The affects of source modelling on the subhalo inference is seldom discussed. In this section, to give a brief idea on how our results would be affected by source modelling, we fit the same mock data with the same mass models discussed above but with a parametric source model. To be specific, the source model we apply here has the same form used to simulated the data, which is an elliptical cored S\'ersic profile. When simulating mock data, we have fixed its break radius to be $0.01\arcsec$, but when using it as a source model, we set its break radius to be a free parameter.

The increase in log likelihood for many model-fits including a subhalo, $\Delta L$, are higher when we assume that the source is an elliptical cored S\'ersic profile as opposed to a pixelised source. This is because the pixelised source models have a much higher level of freedom in how they fit the data. If a mass model provides a good -- but not perfect -- fit, the pixelisation can make small adjustments to the source pixel values to fit the data equally well \citep{Gilman2020b}. This is appropriately penalized using a Bayesian framework (see \citet{Suyu2006} and \citet{Nightingale2018}), but nevertheless produces smaller likelihood contrasts than fitting a parametric source model like the cored S\'ersic profile, which has a lot less freedom in adjusting its parameters in order to account for an inaccurate mass model. This is also dependent on the fact that the elliptical cored S\'ersic profile was used to both simulate and fit the mock strong lenses; had there been a mismatch here parametric fits would likely not give such large $\Delta L$ values.

\begin{figure*}
	\includegraphics[width=2.0\columnwidth]{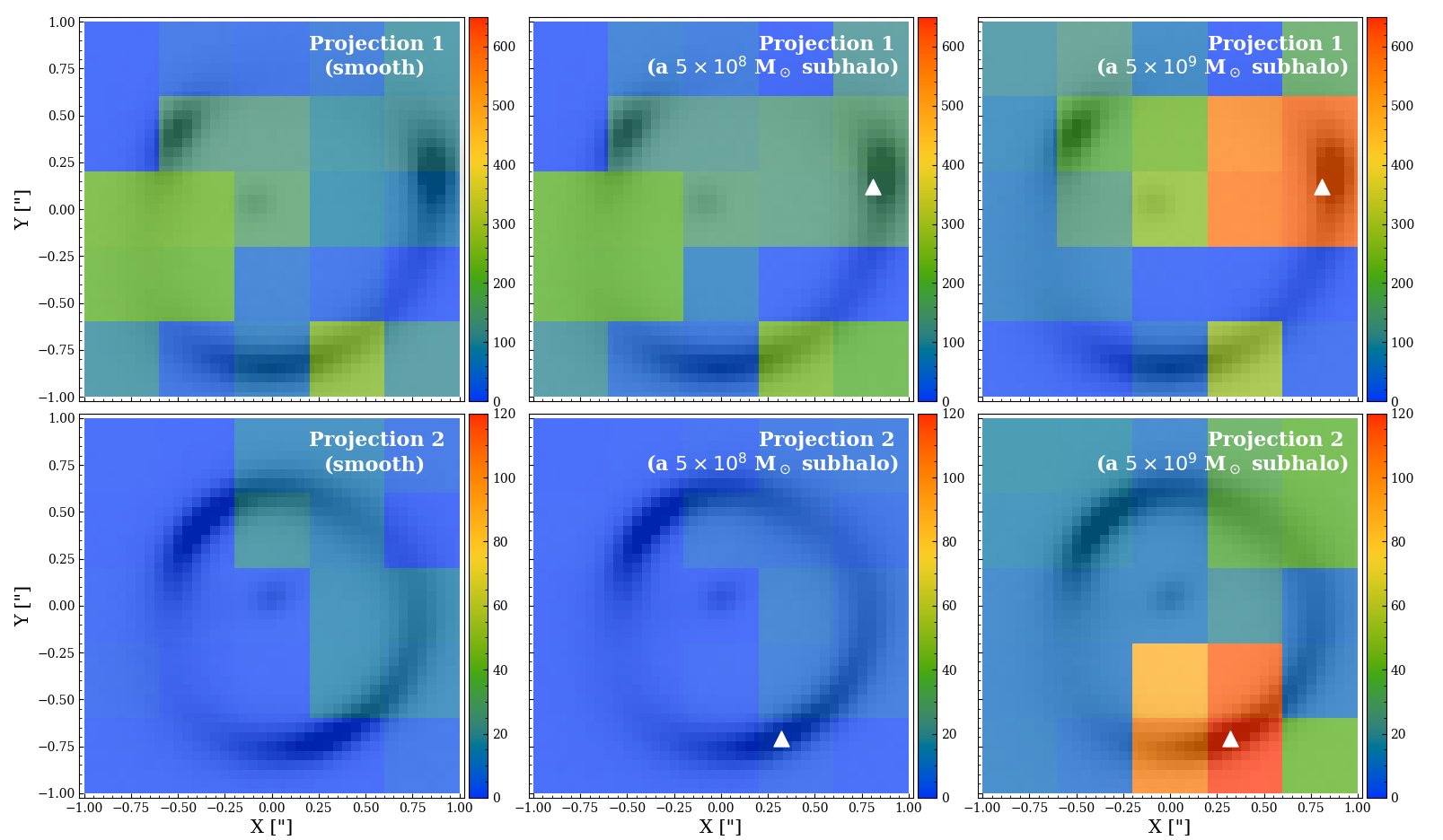}
    \caption{$\Delta L$ maps of using eBPL + cored S\'ersic source model. Colours indicate the increase in maximum log likelihood, $\Delta L$, when a subhalo is included inside $0.4\arcsec\times0.4\arcsec$ squares during a fit to a lens that has: no subhalos (left), a subhalo of mass $5\times10^8$~M$_\odot$ (middle), or a subhalo of mass $5\times10^9$~M$_\odot$ (right). Top and bottom rows show the results for Projection 1 and 2 (with different colour scales). White triangles mark the true locations of the subhaloes}
    \label{fig:lklhd_para_BPL}
\end{figure*}

\begin{figure}
	\includegraphics[width=1.0\columnwidth]{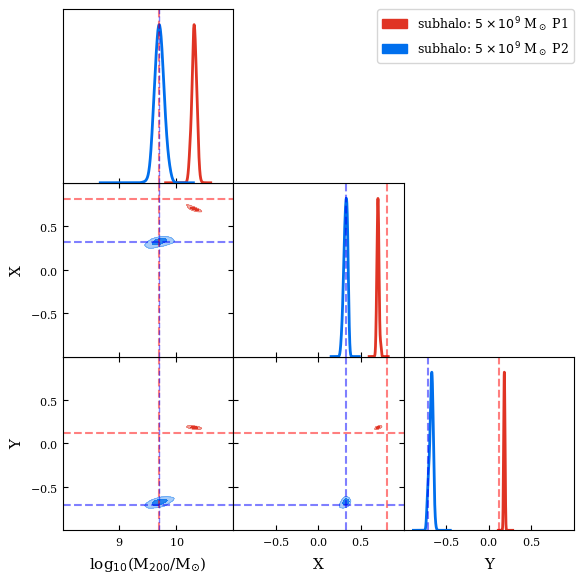}
    \caption{Posteriors of detected subhalo parameters of eBPL + cored S\'ersic model. The red posteriors show the results for an input of a $5\times10^9$~M$_\odot$ subhalo in Projection 1. The blue posteriors show the results for an input of a $5\times10^9$~M$_\odot$ subhalo in Projection 2. The 2D contours cover the 68\% and 99\% confidence regions. The dashed lines in corresponding colors marked the true input values.}
    \label{fig:pos_para_BPL}
\end{figure}

In Fig.~\ref{fig:lklhd_para_BPL}, we show $\Delta L$ when fitting the data with an eBPL profile. For projection 1, for both the smooth case and a $5\times10^8$~M$_\odot$ subhalo input case, the eBPL plus cored S\'ersic source model returns similar results, with a highest $\Delta L$ giving $\sim250$ at the middle left region indicating the existence of a $10^{10.1^{+0.2}_{-0.1}}$~M$_\odot$ subhalo, which is not consistent with our input (e.g. it is a false positive). For the case of a $5\times10^9$~M$_\odot$ input subhalo, the highest $\Delta L$ is $\sim$ 600 around the middle right region, which is consistent with our input. For this case, we further model the subhalo by freeing its position and the posterior we get is shown in color red in Fig.~\ref{fig:pos_para_BPL}. We see that although the position is estimated around the true input, the subhalo's mass is overestimated by around 4 times, which is similar to our previous findings for a pixelised source. 

For projection 2, we see that for the smooth test case, false-positive signals show up in upper right regions with the highest $\Delta L$ to be $\sim18$. Read from the grid of highest $\Delta L$, the best-fit subhalo's mass is $10^{9.3^{+0.4}_{-0.5}}$~M$_\odot$. False positives were not detected for this projection using a pixelised source. For the case of an input subhalo of $5\times10^8$~M$_\odot$, there are some plausible signals around the middle right regions with the highest $\Delta L$ to be $\sim$ 8. The mass of the plausible subhalo obtained in this case is $10^{9.6^{+0.2}_{-0.5}}$~M$_\odot$. For the third case where a $5\times10^9$~M$_\odot$ subhalo added, the $\Delta L$ map returns the correct answer with the highest $\Delta L$ to be 120 at the place where we input the subhalo. For this one, similarly, we further model the subhalo by freeing its position. The posterior is shown in color blue in Fig.~\ref{fig:pos_para_BPL}. We see that the input subhalo can be well recovered in this case. 

\begin{figure*}
	\includegraphics[width=2.0\columnwidth]{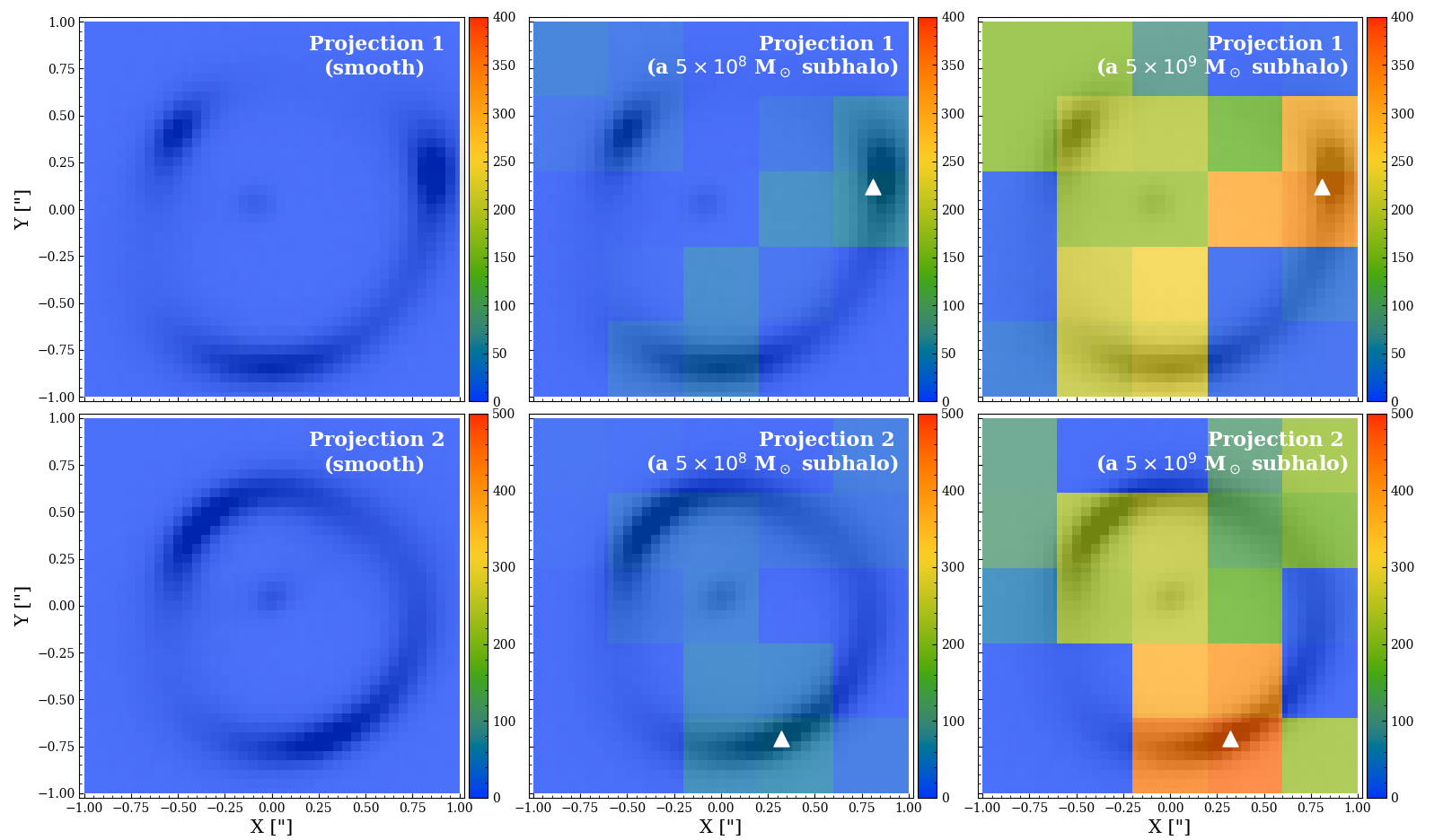}
    \caption{$\Delta L$ maps of using stellar + dark matter + cored S\'ersic source model. Colours indicate the increase in maximum log likelihood, $\Delta L$, when a subhalo is included inside $0.4\arcsec\times0.4\arcsec$ squares during a fit to a lens that has: no subhalos (left), a subhalo of mass $5\times10^8$~M$_\odot$ (middle), or a subhalo of mass $5\times10^9$~M$_\odot$ (right). Top and bottom rows show the results for Projection 1 and 2 (with different colour scales). White triangles mark the true locations of the subhaloes.}
    \label{fig:lklhd_para_3CS}
\end{figure*}

\begin{figure}
	\includegraphics[width=1.0\columnwidth]{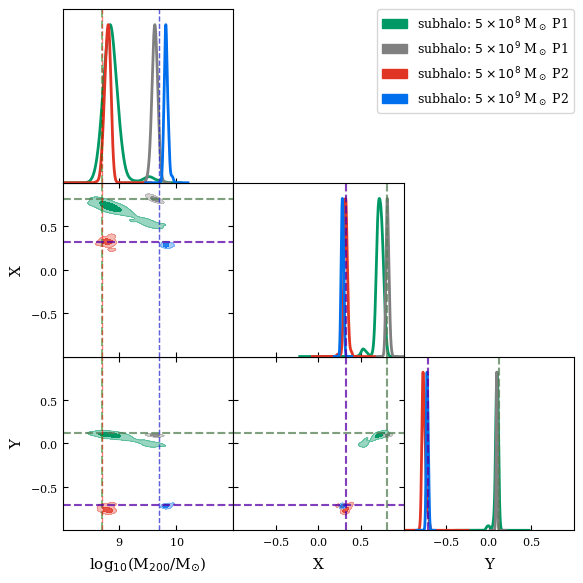}
    \caption{Posteriors of detected subhalo parameters of the stellar + dark matter + cored S\'ersic source model. The green, gray, red and blue posteriors respectively show the results for: an input of a $5\times10^8$~M$_\odot$ subhalo in Projection 1; an input of a $5\times10^9$~M$_\odot$ subhalo in Projection 1; an input of a $5\times10^8$~M$_\odot$ subhalo in Projection 2; an input of a $5\times10^9$~M$_\odot$ subhalo in Projection 2. The 2D contours cover the 68\% and 99\% confidence regions. The dashed lines in corresponding colors marked the true input values.}
    \label{fig:pos_para_3CS}
\end{figure}

In Fig.~\ref{fig:lklhd_para_3CS}, we show $\Delta L$ maps of modelling the data with the decomposed model plus a cored S\'ersic source. We see that the results are similar to the pixelisation results: for smooth tests, no clear false-positive signals show up. For subhalo tests, the highest $\Delta L$ is consistent with the region of an input subhalo. In Fig.~\ref{fig:pos_para_3CS}, we further plot the posteriors obtained for the detected subhaloes. Overall, input subhaloes can be recovered to a good level although for the $5\times10^9$~M$_\odot$ subhalo cases, the recovered masses are slightly offset to the true value, albeit this is close enough that it could simply be due to noise in the mock observation. 

\bsp	
\label{lastpage}
\end{document}